\documentclass[prd,nofootinbib,preprint]{revtex4}
\UseRawInputEncoding
\usepackage{graphicx}
\usepackage{url}
\usepackage{tikz}
\usepackage[bf]{caption}
\usepackage{amsmath}
\usepackage{cases} 
\usepackage{multirow}

\usepackage{epsfig}
\usepackage{makecell}
\usepackage{amsmath}
\usepackage{amsfonts}
\usepackage{amssymb}
\usepackage{mathtools}
\usepackage{url}
\usepackage{subfigure}
\usepackage{hhline}
\usepackage{color}
\usepackage{array,multirow}
\usepackage{bm}
\usepackage{makecell}
\usepackage{epsfig}
\usepackage{amsmath}
\usepackage{amsfonts}
\usepackage{amssymb}
\usepackage{url}
\usepackage{hyperref}
\usepackage{subfigure}
\usepackage{hhline}
\usepackage{color}
\usepackage{bm}
\usepackage{cancel}
\usepackage{mathtools}

\newcommand{\beqa}{\begin{eqnarray}}
\newcommand{\eeqa}{\end{eqnarray}}
\newcommand{\beq}{\begin{equation}}
\newcommand{\eeq}{\end{equation}}

\newcommand{\nn}{\nonumber}
\newcommand{\bmt}{\begin{pmatrix}}
\newcommand{\emt}{\end{pmatrix}}
\usepackage[toc,page]{appendix}
\usepackage{comment}
\newcommand{\be}{\begin{equation}}
\newcommand{\ee}{\end{equation}}
\newcommand{\bea}{\begin{eqnarray}}
\newcommand{\eea}{\end{eqnarray}}

\usepackage{epsfig}  
\usepackage{graphicx}
\usepackage{hyperref}

\hypersetup{
    colorlinks=true,
    citecolor=green,
    linkcolor=magenta,
    filecolor=green,      
    urlcolor=magenta,
}
\usepackage{tabularx}
\usepackage{color}
\usepackage{float}
\usepackage{amsfonts}
\usepackage{amsmath}
\usepackage{slashed}
\usepackage{soul}
\usepackage{comment}
\def\beq{\begin{equation}}
\def\eeq{\end{equation}}
\def\bea{\begin{eqnarray}}
\def\eea{\end{eqnarray}}
\def\nn {\nonumber}
\newcommand{\bpl}{\beta_{+}}
\newcommand{\bmi}{\beta_{-}}
\newcommand{\mm}{m_{-}}
\newcommand{\mpl}{m_{+}}
\newcommand{\qsq}{q^{2}}
\def\re{{\rm Re}}

\large
\begin{document}
\title{
Delving into the $ B_s \to \ell \ell^{\prime}$, $B_{(s)} \to (K^{(*)}, \phi, f_2^{\prime}, K_2^*) \ell \ell ^{\prime}$ processes
}

\author{Manas K. Mohapatra}
\email{manasmohapatra12@gmail.com}

\author{Lopamudra Nayak}
\email{lopamudra.nayakcda@gmail.com}

\author{Rashmi Dhamija}
\email{ph19resch01003@iith.ac.in}

\author{Anjan Giri}
\email{giria@iith.ac.in}



         
\affiliation{Department of Physics, IIT Hyderabad,
              Kandi - 502285, India }                            


\begin{abstract}
To shed light on the indirect search for new physics beyond the standard model, the long standing discrepancies between the theory and experiment mediated by FCNC $b\to s \ell \ell$ quark level transitions set an ideal testing ground. Though the very recent measurements of $R_K$ and $R_{K^*}$ are consistent with the standard model, still the excitements remain on the measurements of LHCb experiment with the observables $\mathcal{B} (B_s \to \phi \mu ^+ \mu ^-)$ which has deviations at the level of $3.6 \sigma$. Additionally, standard deviation of $\sim 3.3 \sigma$ and $1.2 \sigma$, respectively for $P_5^{\prime}$ in $B \to K^* \mu ^+ \mu ^-$ and the branching ratio in $B_s \to \mu^+ \mu^-$ processes are observed. Inspired by these discrepancies, we work out the constraints on the new physics coupling parameters in the presence of a non-universal $Z'$ model. We then probe the exclusive leptonic decay channels $ B_s \to \ell \ell^{\prime}$, $B_{(s)} \to (K^{(*)}, \phi, f_2^{\prime}, K_2^*) \ell \ell ^{\prime}$ induced by the neutral current transition $b\to s \ell \ell^{\prime}$. We find that the $q^2$ variation of the observables, such as,  branching ratio, forward-backward asymmetry, lepton polarization asymmetry, and the very sensible observable, so called non-universality observables for LFV decays display the sensitivity of new physics. In this analysis. we estimate above mentioned observables that could shed light on the window of new physics in the near future.
\end{abstract}
\maketitle


\section{Introduction}
Our best understanding of how the particles and three of the forces such as strong, electromagnetic and weak interactions communicate to each other is encapsulated in the Standard Model (SM) of particle physics. 
Over time and through many experiments, the SM has proven a well established physics theory.
Despite its spectacular success at explaining the data, the SM is believed to be incomplete which fails to describe a few challenging shortcomings such as the matter dominance over anti-matter in the present universe, neutrino masses, hierarchy problem, dark matter and dark energy, the unification of gravity
with other three fundamental forces etc. To solve  these open puzzles, the quest for physics beyond the SM (BSM) is of prime importance. In this context, the B factories have been an excellent testing ground to shed light in exploring the new physics (NP) beyond the SM through low energy experiments. It has literally witnessed the breaking of lepton flavor universality in the charged (neutral) current decays mediated by $b \to c$ ($b \to s$) quark level transitions. Experimentally, the $b \to c \tau \nu$ quark level transition have lepton non universality anomalies in exclusive  $B \to D^{(*)}\tau \nu$, $B \to J/\psi \tau \nu$ decays with a tension of $1.4 \sigma$ ($2.5 \sigma$) and $1.8 \sigma$, respectively, obtained from the HFLAV group~\cite{HFLAV:2022pwe}. The $\tau$ polarization fraction and the longitudinal polarization fraction of $D^*$ meson in $B \to D^* \tau \nu$ have $1.6 \sigma$~\cite{Belle:2016dyj, Belle:2017ilt} and $1.6 \sigma$~\cite{Belle:2019ewo} deviations, respectively. Similarly, several measurements in $b \to s \ell ^+ \ell ^-$ transitions such as the angular observable $P^{\prime}_5$ in $B \to K^{*} \mu ^+ \mu ^-$ in the bins
$q^2 \in[4.0, 6.0]$, [4.3, 6.0] and [4.0, 8.0] from ATLAS~\cite{ATLAS:2018gqc}, LHCb~\cite{LHCb:2013ghj,LHCb:2015svh}, CMS~\cite{CMS:2017rzx}, 
Belle~\cite{Belle:2016xuo}
respectively deviate at $3.3\sigma$, $1\sigma$ and $2.1\sigma$ from the SM expectations~\cite{Descotes-Genon:2012isb,Descotes-Genon:2013vna,Descotes-Genon:2014uoa}. 
The updates in the measurement of the the branching fraction of $B_s \to \phi \mu ^+ \mu ^-$~\cite{LHCb:2021zwz,LHCb:2013tgx,LHCb:2015wdu} 
in $q^2\in[1.1, 6.0]$ region has discrepancy at the level of $3.6\sigma$ from the SM expectations~\cite{Aebischer:2018iyb,Bharucha:2015bzk} . However, the recent updates by LHCb Collaboration~\cite{LHCb, LHCb1} in the measurements of 
$R_K=\mathcal{B}(B \to K \mu ^+ \mu^-)/\mathcal{B}(B \to K e ^+e ^-)$ and $R_{K^*}=\mathcal{B}(B \to  K^{*}\, \mu^+ \mu ^-)/\mathcal{B}(B \to K^{*} e ^+ e ^-)$,
in the bin range $q^2\in [0.1, 1.1]$ and $[1.1, 6.0]$, are consistent with the SM predictions. On the other hand, the lepton flavor violating (LFV) decays are forbidden at the tree-level in the SM and can in principle occur via neutrino mixing through loop and box diagrams. Because of such mixing, the rate is considerably below current or future experimental sensitivities. Consequently, this causes mixing between different generations of leptons which give rise to flavor-changing neutral current (FCNC) transitions. Keeping this in mind, the leptonic LFV decays such as $\tau \to \mu \mu \mu,$ $\tau \to eee$, $\mu \to eee$, etc have been analysed in various NP models though the experimental upper limit exists~\cite{ParticleDataGroup:2022pth}. However, the principle in the FCNC transition of the quark sector for LFV decays could be similar to that of the lepton sector. In this regard, we explore the exclusive LFV decays in quark sector such as $B_s  \to \ell \ell ^{\prime}$, and $B_{(s)} \to (K^{(*)}, \phi, f_2^{\prime}, K_2^*) \ell \ell ^{\prime}$ decays which occur via $b \to s \ell \ell ^{\prime}$ quark level transition. 
Experimentally, these decay channels are not yet observed but the upper limits of few observables exist. The leptonic $B_s \to \mu e$ and $B_s \to \tau \mu$ processes have upper limits of $5.4 \times 10^{-9}$ and $4.2 \times 10^{-5}$, respectively by LHCb Collaboration~\cite{LHCb:2017hag, LHCb:2019ujz}. Similarly, the upper bounds are measured by LHCb and BaBar collaboration in the  branching ratios of $B \to K \ell \ell ^{\prime}$ processes are $\mathcal{B}(B \to K e \mu)<7.0 \times 10^{-9}$~\cite{LHCb:2019bix}, $\mathcal{B}(B \to K \tau \mu)<$ $1.5 \times 10^{-5}$~\cite{BaBar:2012azg}, and $\mathcal{B}(B \to K e \tau)<4.5 \times 10^{-5}$~\cite{BaBar:2012azg}, respectively. The upper limit of the branching ratio of $B^0 \to K^* e \mu$ channel is observed as $1.8 \times 10^{-7}$ by Belle Collaboration~\cite{Belle:2018peg}. We analyse the $B \to K_2^* \ell \ell ^{\prime}$ process because a better understanding of $B \to K_2^* \gamma$ channel has been given by BaBar Collaboration in ref.~\cite{BaBar:2003aji}. On the other hand, the $B \to f_2 ^{\prime} \ell \ell ^{\prime}$ process provides very less attention both in theory and experiment. Thus , it can be studied similarly to the $B \to K_2 ^* \ell \ell ^{\prime}$ process.
It is interesting to see if the associated observables could be enhanced in some new physics models that could simultaneously explain the observed $b\to s \ell \ell$ data. 
In this analysis, we consider a simplified non-universal $Z^{\prime}$ model in which the NP effects originate from $U(1)^{\prime}$ abelian group extension to the SM gauge symmetry. Consequently, it provides a heavy new gauge boson $Z'$ of mass $m_{Z^{\prime}}$ with generic couplings to quarks and leptons, and induces FCNC transition at tree level. Inspired by these available upper limits, we study the above discussed LFV decays in the presence of non-universal $Z^{\prime}$ model.

The outline of the paper is as follows. In section \ref{TT}, we discuss the theoretical toolkit that includes the most general effective weak Hamiltonian for $b \to s \ell \ell ^{\prime}$ NP operators. We also report the relevant formula for all the decay observables pertaining to $B_s \to \ell \ell ^{\prime}$, $B_{(s)} \to (K^{(*)}, \phi, f_2^{\prime}, K_2^*) \ell \ell ^{\prime}$ decay channels. In section \ref{NP}, we discuss the new physics analysis in the presence of non-universal $Z'$ model by using the updated experimental limits on the $b \to s \ell \ell$ data. In section \ref{NAD}, we discuss the numerical analysis of the aforementioned observables of rare (semi)leptonic LFV decays. We conclude with the summary of
our results in section \ref{CON}.


\newpage
\section{Theoretical Toolkit:}\label{TT}
\subsection{Effective Hamiltonian}
In this section, we focus on the exclusive lepton flavor violating $b \to s \ell \ell^{\prime}$ $(\ell, \ell ^{\prime} =e, \mu , \tau$) transition processes. In the SM, the leptons $\ell$ and $\ell^{\prime}$ are considered to have same flavor whereas the non-universal $Z^{\prime}$ boson couple differently in the NP models. The most structured weak effective Hamiltonian describing the $b\to s\ell \ell^{\prime}$ processes can be represented as~\cite{Dedes:2008iw, Becirevic:2016zri, Crivellin:2015era},

\bea
\mathcal{H}^{eff} =- \frac{G_F \alpha}{2\sqrt{2}\pi}  V_{tb} V_{ts}^* \sum _{m=9,10} C_m^{NP}O_m +h.c.,
\eea \label{effH}
where $G_F$ ($\alpha$) represents the Fermi (electromagnetic) coupling constant and $V_{tb}V_{ts}^*$ is the CKM matrix element.  The primed counter parts of the operators can be obtained by replacing $P_L \rightleftharpoons P_R$. It is very sensitive to the semileptonic operators $O_9$ and $O_{10}$ and are given by,

\bea
O_9 = [\bar{s}\gamma _ \mu (1-\gamma _5)b] [\bar{\ell \gamma ^ \mu \ell ^{\prime}}], \hspace{1cm}
O_{10} = [\bar{s}\gamma _ \mu (1-\gamma _5)b] [\bar{\ell \gamma ^ \mu \gamma _5 \ell ^{\prime}}].
\eea


The standard decomposition of the hadronic matrix element are given as,

\bea
\langle 0|\bar{b}\gamma P_L(R)s|B_s(p)\rangle = \pm \frac{i}{2}p_\mu f_{B_s}, \nn\\
\langle 0|\bar{b}\gamma P_L(R)s|B_s(p)\rangle = \pm \frac{i}{2}p_\mu f_{B_s}, \nn\\
\langle 0|\bar{b}P_L(R)s|B_s(p)\rangle = \pm \frac{i}{2}\frac{M_{B_s}^2 f_{B_s}}{m_b+m_s}, \nn\\
\eea

where $f_{B_s}$  and p$_\mu$ are the decay constant   and momentum of the $B_s$ meson, respectively.

\subsection{Decay observables of (semi)leptonic LFV $b \to s \ell \ell ^{\prime}$ processes:} 

\subsubsection{$B_s \to \ell \ell ^{\prime}$}
From the effective Hamiltonian (\ref{effH}) one can obtain the amplitude and the associated branching ratios of $B_s \rightarrow \ell \ell^{\prime}$ process in the SM are given as \cite{Sheng:2018qtp, Duraisamy:2016gsd,Sheng:2018ylg, Becirevic:2016zri};

\begin{align}
\label{Bsformula}
&
{\cal B}(B_s\to \ell \ell ^{\prime}) =\dfrac{\tau_{B_s}}{64 \pi^3}\frac{\alpha^2 G_F^2}{m_{B_s}^3}  f_{B_s}^2 |V_{tb}V_{ts}^*|^2\lambda^{1/2}(m_{B_s},m_\ell,m_{\ell^{\prime}})\nn \\
&\qquad\times \Bigg{\lbrace}[m_{B_s}^2-(m_\ell+m_{\ell^{\prime}})^2]\cdot\left|(C_9^{NP}-C_9')(m_\ell-m_{\ell^{\prime}})\right|^2\nn\\
&\hspace{1cm}+[m_{B_s}^2-(m_\ell-m_{\ell^{\prime}})^2]\cdot\left|(C_{10}^{NP}-C_{10}')(m_\ell+m_{\ell^{\prime}})\right|^2\Bigg{\rbrace},
\end{align}
where  $\lambda(a,b,c)=[a^2-(b-c)^2][a^2-(b+c)^2]$. 

\subsubsection{$B\to K \ell \ell ^ \prime$}

The semileptonic $B \to K \ell \ell^ \prime$ decay mode involves $b \to s$ quark level transition mediated by the effective Hamiltonian (\ref{effH}). Here the kinematic variables given in Ref.~\cite{Korner:1989qb} are defined in such a way that the main decay axis, denoted by $z$, is defined in the rest frame of $B$ meson whereas the K meson, and the lepton pair $\ell$ and $\ell ^ {\prime}$ travel in the opposite directions. The polar angle $\theta_\ell$ is the angle between the meson K and the lepton $\ell$ in the $\ell - \ell ^ \prime$ rest frame.

The standard parametrizations for hadronic matrix elements are provided by
\begin{align}
\langle \bar K(k)|\bar{s}\gamma_\mu b|\bar B(p)\rangle &= \Big{[}(p+k)_\mu- \frac{m_B^2-m_K^2}{q^2}q_\mu \Big{]}f_+(q^2)+\frac{m_B^2-m_K^2}{q^2} q_\mu f_0(q^2),\\
\langle \bar K(k)|\bar{s}\sigma_{\mu\nu} b|\bar B(p)\rangle &= -i (p_\mu k_\nu-p_\nu k_\mu)\frac{2 f_T(q^2,\mu)}{m_B+m_K}.
\end{align}
The hadronic form factors (FFs) $f_{+}(q^2)$, $f_{0}(q^2)$ and $f_{T}(q^2)$ are the functions of $q^2$ which lies between $(m_1 +m_2)^2 $ and $ (m_B-m_K)^2$.
By employing the above definitions, the differential decay rate can be written as~\cite{Sheng:2018qtp, Duraisamy:2016gsd,Sheng:2018ylg, Becirevic:2016zri},
\begin{align}
\label{eq:semilepPP}
\dfrac{\mathrm{d}{\cal B}}{\mathrm{d}q^2}(\bar B \to \bar K \ell_1^- \ell_2^+)& = \vert{\cal M}_{K}(q^2)\vert^2\times\Big\lbrace 
\varphi_7(q^2) |C_7+C_{7}'|^2 + \varphi_{9}(q^2) |C_{9}+C_{9}'|^2  +  \varphi_{10}(q^2) |C_{10} +C_{10}'|^2   \nonumber \\
& + \varphi_S(q^2) |C_S+C_{S}'|^2+ \varphi_P(q^2) |C_P+C_{P}'|^2 + \varphi_{79}(q^2) \mathrm{Re}[(C_7+ C_7^\prime) (C_{9}+ C_9^\prime)^*]  \nonumber \\
&+ \varphi_{9S}(q^2) \mathrm{Re}[(C_{9}+ C_9^\prime) (C_S+ C_S^\prime)^*]+ \varphi_{10P}(q^2) \mathrm{Re}[(C_{10} +C_{10} ^\prime)(C_P+C_P^\prime)^*]  \Big\rbrace ,
\end{align}
where $\varphi_{i}(q^2)$ depends on kinematical quantities and on the form factors and shown in Appendix~\ref{BtoKparameters}.
The normalization factor given in eq.~(\ref{eq:semilepPP}) reads
\begin{equation}
\vert {\cal M}_{K}(q^2)\vert^2=\tau_{B_d}\dfrac{\alpha^2 G_F^2 |V_{tb}V_{ts}^*|^2}{512 \pi^5 m_B^3}\dfrac{\lambda^{1/2}(\sqrt{q^2},m_1,m_2)}{q^2}\lambda^{1/2}(\sqrt{q^2},m_B,m_K),
\end{equation}
whereas the kinematic factor is given as 
\bea
\lambda=m_B^4+m_K^4+q^4-2(m_B^2m_K^2+mK^2q^2+m_B^2q^2).\label{knfactor}\eea 

\subsubsection{$B\to K^\ast \ell \ell ^{\prime}$ and $B_s\to \phi \ell \ell ^{\prime}$}
\begin{figure}
\includegraphics[scale=0.25]{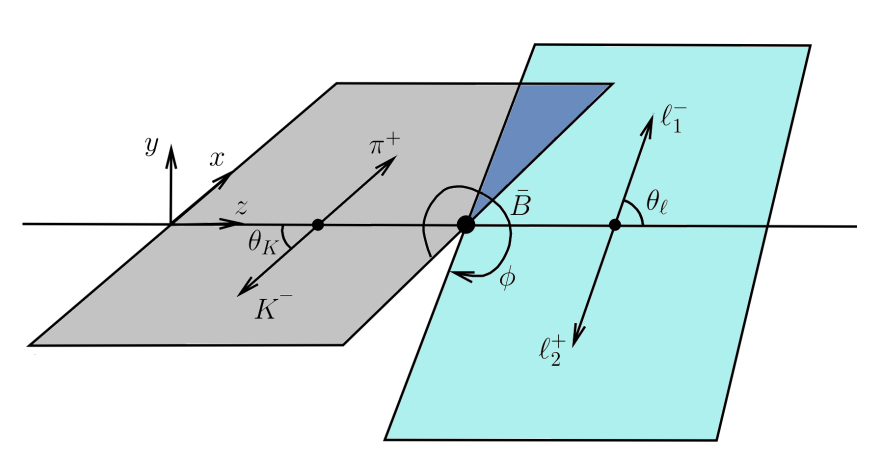} 
\caption{Angular distribution of $B \to K^* (\to K \pi) \ell \ell ^ {\prime}$ processes}
\label{Angdist}
\end{figure}

Here we focus on $B \to V \ell \ell ^{\prime}$ ($V= K^*, \phi$) decays proceeding via $b \to s \ell \ell ^{\prime}$ processes where the vector mesons further decay as $K^* \to K \pi$ and $\phi \to K\bar{K}$, respectively. We also express the angular distributions of $B \to K^*(\to K \pi) \ell \ell ^{\prime}$ process. Similarly, the distributions associated with the $B_s \to \phi$ transition can be obtained by trivial replacement of the form factor and the mass of the particle involved. We adopt the details concerning the kinematics from Ref.~\cite{Korner:1989qb}. In the angular conventions~\cite{Sheng:2018qtp, Duraisamy:2016gsd,Sheng:2018ylg,Becirevic:2016zri}, $\theta _ \ell$ is the angle between the lepton $\ell$  with the decay axis in the lepton pair rest frame, $\theta _K$ is the made by the decay axis with the direction of flight of $K$ meson in the rest frame of $K^*$ vector meson. The angle $\phi$ is the angle spanned between the $K \pi$ and $\ell \ell ^ {\prime}$ planes, respectively shown in Fig.~\ref{Angdist}.
The transition amplitude of exclusive $B\to K^* \ell \ell ^{\prime}$  decay mode are associated with the hadronic matrix elements. These are parametrized in terms of form factors, and are given as~\cite{Becirevic:2016zri}

\begin{align}\label{def:FFV}
\langle \bar{K}^\ast(k)|\bar{s}\gamma^\mu(1-\gamma_5) b|\bar{B}(p)\rangle &= \varepsilon_{\mu\nu\rho\sigma}\varepsilon^{\ast\nu}p^\rho k^\sigma \frac{2 V(q^2)}{m_B+m_{K^\ast}}-i \varepsilon_\mu^\ast(m_B+m_{K^\ast})A_1(q^2)\\[.3em] 
&+i(p+k)_\mu (\varepsilon^\ast \cdot q)\frac{A_2(q^2)}{m_B+m_{K^\ast}}+i q_\mu(\varepsilon^\ast \cdot q) \frac{2 m_{K^\ast}}{q^2}[A_3(q^2)-A_0(q^2)],\nonumber\\[.7em] 
\langle \bar{K}^\ast(k)|\bar{s}\sigma_{\mu\nu} q^\nu(1-\gamma_5) b|\bar{B}(p)\rangle &= 2 i \varepsilon_{\mu\nu\rho\sigma} \varepsilon^{\ast\nu}p^\rho k^\sigma T_1(q^2)+[\varepsilon_\mu^\ast(m_B^2-m_{K^\ast}^2)-(\varepsilon^\ast \cdot q)(2p-q)_\mu]T_2(q^2)\nonumber\\[.3em] 
&+(\varepsilon^\ast \cdot q)\Big{[}q_\mu - \frac{q^2}{m_B^2-m_{K^\ast}^2}(p+k)_\mu \Big{]}T_3(q^2),
\end{align}
where $\varepsilon_\mu$ is the polarization vector of $K^\ast$ meson. The transition form factor $A_3(q^2)$ is associated to the combinations of both $A_{1}(q^2)$ and $A_{2}(q^2)$ form factors and is given as 
\bea 2 m_{K^\ast} A_3(q^2)=(m_B+m_{K^\ast})A_1(q^2)-(m_B-m_{K^\ast})A_2(q^2).
\eea

The full angular distribution of the $B \to K^* \ell \ell ^\prime$ decay mode can be read as~ 
\begin{equation}
\dfrac{\mathrm{d}^4 {\cal B} ({B}\to {K}^{\ast}(\to K\pi) \ell \ell ^{\prime})}{\mathrm{d}q^2\mathrm{d}\cos \theta_\ell \mathrm{d}\cos \theta_K \mathrm{d}\phi} = \dfrac{9}{32\pi}I(q^2,\theta_\ell,\theta_K,\phi),
\end{equation}
where
\begin{align}
I(q^2,\theta_\ell,\theta_K,\phi) = &I_1^s(q^2)\sin^2\theta_K + I_1^c(q^2)\cos^2\theta_K+[I_2^s(q^2)\sin^2\theta_K+I_2^c(q^2)\cos^2\theta_K]\cos 2\theta_\ell\nn\\[.4em] 
&+I_3(q^2)\sin^2\theta_K \sin^2\theta_\ell \cos 2\phi+I_4(q^2)\sin 2\theta_K \sin 2\theta_\ell \cos \phi \nn \\[.4em] 
&+ I_5(q^2) \sin 2\theta_K\sin \theta_\ell\cos\phi+[I_6^s(q^2)\sin^2\theta_K+I_6^c(q^2)\cos^2\theta_K]\cos \theta_\ell \nonumber \\[.4em] 
&+I_7(q^2)\sin 2\theta_K \sin \theta_\ell \sin \phi + I_8(q^2)\sin 2\theta_K \sin 2\theta_\ell \sin\phi \nonumber\\[.4em] 
&+I_9(q^2) \sin^2\theta_K \sin^2\theta_\ell \sin 2 \phi.
\end{align}
The $q^2$-dependent differential branching fraction, after integrating over the physical region of the phase space $\theta _K$, $\theta _ \ell$ and $\phi$, is simply given as
\bea
{\mathrm{d}{\cal B}\over \mathrm{d}q^2}=\frac{1}{4}\left[3 I_1^c(q^2)+6 I_1^s(q^2)-I_2^c(q^2)-2I_2^s(q^2)\right]\,
\eea 
 with
\bea
 (m_i+m_j)^2 \leq q^2 \leq (M_B-M_{K^*})^2, \hspace{0.25cm} -1 \leq \cos\theta_l \leq 1, \hspace{0.25 cm}
 -1 \leq \cos\theta_K \leq 1, 
\hspace{0.25cm} 0 \leq \phi \leq 2\pi.
\eea

Here the angular coefficients $I^i_j(q^2)$ (i= c,s ; j= 1, 2,..9) are defined in terms of the transversity amplitudes
and given in Appendix~\ref{BtoKstarparameters}. 

\subsubsection{$B \to T \{K_2 ^* ,f_2 ^{\prime}\} \ell \ell ^{\prime}$ processes}
In contrast to the previous sub-section, we study the exclusive $B\to T~(K_2^*, f_2 ^{\prime}) \ell \ell^{\prime}$  decay modes mediated via $b\to s$ quark level transition.
The long distance contribution in terms of the hadronic matrix element of $B\to K_2^*$ transition are given as~\cite{Wang:2010ni, Yang:2010qd}
 \begin{eqnarray}
  \langle K_2^*(k, \epsilon ^*)|\bar s\gamma^{\mu}b|\overline B(p)\rangle
  &=&-\frac{2V(q^2)}{m_B+m_{K_2^*}}\epsilon^{\mu\nu\rho\sigma} \epsilon ^*_{T\nu}  p_{\rho}k_{\sigma}, \nonumber\\
  \langle  K_2^*(k,\epsilon ^*)|\bar s\gamma^{\mu}\gamma_5 b|\overline B(p)\rangle
   &=&2im_{K_2^*} A_0(q^2)\frac{\epsilon^*_{T } \cdot  q }{ q^2}q^{\mu} + i(m_B+m_{K_2^*})A_1(q^2)\left[ \epsilon^{*\mu}_{T}
    -\frac{\epsilon ^*_{T } \cdot  q }{q^2}q^{\mu} \right] \nonumber\\
    &&-iA_2(q^2)\frac{\epsilon ^*_{T} \cdot  q }{  m_B+m_{K_2^*} }
     \left[ (p+k)^{\mu}-\frac{m_B^2-m_{K_2^*}^2}{q^2}q^{\mu} \right],
\end{eqnarray}
where $p$ ($k$) is the four momentum of $B$ ( $K_2^{*}$) meson.
We use the relevant form factors in our analysis for $B_{(s)}$ to light $J^{PC}=2^{++}$ tensor meson ($T$) derived from the light-cone sum rule (LCSR) approach. 
Within this technique, the parameterized $q^2$ dependent form factors are given in the form as \cite{Yang:2010qd}:
\begin{equation}
 F^{B_{(s)}T}(q^2)=\frac{F^{B_{(s)}T}(0)}{1-a_T(q^2/m_{B_q}^2)+b_T(q^2/m_{B_q}^2)^2},
\end{equation}
where $F=V,A_{0,1,2}$ and $T_{1,2,3}$ are the transition form factors.

The $B \to K_2^* \ell \ell^{\prime}$ decay mode which undergoes $b\to s \ell \ell^{\prime}$ quark level transition can be expressed in terms of the leptonic polar angle $\theta_\ell$ and leptonic mass squared $q^2$. The angle $\theta_\ell$ is the angle made by the lepton $\ell$ with respect to the di-lepton momentum in the rest frame of $B$ meson. The two-fold differential decay distribution in terms of the variables $\theta _\ell$ and $q^2$ is given as follows~\cite{Kumbhakar:2022szr}
\begin{equation}
    \frac{d^2\Gamma}{dq^2 d \cos \theta_\ell}=A(q^2) + B(q^2) \cos \theta_\ell + C(q^2) \cos ^2\theta_\ell,
    \label{dist}
\end{equation}
where the $q^2$ parameters $A(q^2)$, $B(q^2)$ and $C(q^2)$ includes form factors and Wilson coefficients. The detailed expressions are given in Appendix~\ref{appenb}. 
 Now after integrating Eq.~(\ref{dist}) over $\theta_\ell$, we obtain the differential branching ratio  as
\begin{equation}
    \frac{d\mathcal{B}}{d q^2}= 2\tau_B \left(A  + \frac{C}{3}\right),
\end{equation}
and the lepton forward-backward asymmetry is represented as
\begin{equation}
A_{\rm FB}(q^2)= \frac{1}{d\Gamma/dq^2}\left(\int_0^1 d\cos\theta_\ell\frac{d\Gamma}{d\cos\theta_\ell d q^2}-\int_{-1}^0 d\cos\theta_\ell \frac{d\Gamma}{d\cos\theta_\ell d q^2 }\right) = \frac{B}{2\left(A+\frac{C}{3}\right)}.
\end{equation}
Similarly, one can do the analysis for $B \to f_2^{\prime} \ell \ell$ process like the $B\to K_2^{*}$ transition where the form factor can be obtained from Ref.~\cite{Yang:2010qd}. 
Analogously, we  would like to see whether it is possible to observe non-universality  in  the  LFV decays. Hence, we define the ratios of branching ratios of  various LFV $b \to s \ell \ell^{\prime}$ decays as
\bea
&&R_{K \ell}^{\ell \ell ^{\prime}} = \frac{{ \mathcal{B}}\left( \bar{B} \to \bar{K } \ell \ell ^{\prime} \right)}{{\mathcal{B}}\left( \bar{B} \to \bar{K } \ell \ell \right)},\\
&&R_{V\ell}^{\ell \ell ^{\prime}} = \frac{{\mathcal{B}}\left( \bar{B} \to \bar{V } \ell \ell ^{\prime} \right)}{{\mathcal{B}}\left( \bar{B} \to \bar{V } \ell \ell \right)}, (V=K^* , \phi), \\
&&R_{Kl}^{\ell \ell ^{\prime}} = \frac{{\mathcal{B}}\left( \bar{B} \to \bar{ T} \ell \ell ^{\prime} \right)}{{\mathcal{B}}\left( \bar{B} \to \bar{T } \ell \ell \right)}, (T= K_2^*, f_2^{\prime}),
\eea \label{LNU}
with $\ell= e, \mu$.
\section{New Physics Analysis in the non-universal $Z^{\prime}$ model:}\label{NP}
Out of all the physics beyond the SM scenarios, an extra abelian $U(1)^{\prime}$ gauge group in extension to the SM is more ubiquitous, which provides a neutral massive vector (spin - 1)  boson $Z^{\prime}$ \cite{Langacker:2000ju}. 
At tree level, this heavy $Z^{\prime}$ boson does flavor changing neutral current transition proceeding via $b\to s(d) \ell \ell^{\prime}$ quark level. It is the most obvious candidate that can evolve in the form of weak effective Hamiltonian of $b\to s(d)$ quark level transition. This is also responsible for an appreciable deviation from the SM results and explain the collider data. 

\begin{figure}[htb]
\includegraphics[scale=0.35]{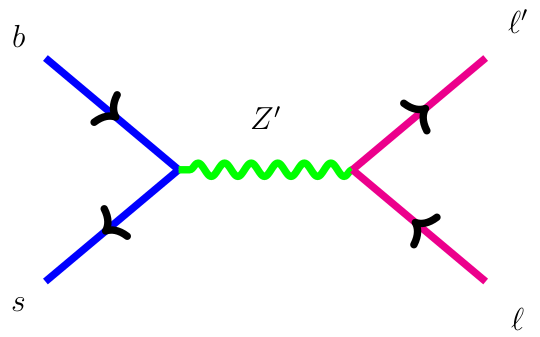} 
\caption{Tree level exchange of $Z^{\prime}$ boson for $b\to s \ell \ell ^{\prime}$ process.}
\label{Feynmandiag}
\end{figure}
In this work, we will formalise it's application to the case of $b \to s$ transition (In general, it is straightforward to generalize for $b \to d$ quark level transition).
There are different kinds of new physics models, such as $Z^{\prime}$, leptoquark (LQ), FCNC mediated $Z$ boson model etc, and have been analysed in Refs~\cite{Mohapatra:2021ynn, Mohapatra:2021izl, Mohapatra:2019wcm, Mohapatra:2021zdp, Mohanta:2010yj, Duraisamy:2016gsd, Sahoo:2015wya}. With the tree level exchange, the new physics scenarios for the parton level $b\to s \ell \ell ^{\prime}$ can be explained in two scenarios: $\mathcal{S}$ - ${ \rm I}:C_9^ {NP}\neq 0$  and $\mathcal{S}$ - $ { \rm II}:C_9^ {NP} = - C_{10}^ {NP}$  \cite{Crivellin:2015era,Capdevila:2017bsm}. However, these two scenarios are possible in $Z^{\prime}$ model whereas LQ does only scenario ($\rm II$). In this analysis, we probe the $b\to s \ell \ell ^{\prime}$ exclusive decays in the presence of non-universal $Z^{\prime}$ model. The Feynman diagram in the presence of $Z^{\prime}$ boson for the exclusive $b\to s \ell \ell ^{\prime}$ decays is given in Fig.~\ref{Feynmandiag}.
The new physics couplings associated with the $Z^{\prime}$ model can be read as~\cite{Biswas:2022wcz, Crivellin:2015era}
\bea
C_{9,10}^{NP} = -\frac{\pi}{\sqrt{2} M_{Z^{\prime}}^2}\frac{1}{\alpha G_F V_{tb}V_{ts}^*}\Gamma _{bs}^{L}(\Gamma _{\ell \ell ^{\prime}}^R\pm \Gamma _{\ell \ell ^{\prime}}^L) .
\eea
The scenario $(\rm I)$ can be obtained with $\Gamma _{\ell \ell ^{\prime}}^L= \Gamma _{\ell \ell ^{\prime}}^R$, whereas the scenario $\rm (II)$ comes with the condition $\Gamma _{\ell \ell ^{\prime}}^R=0$. Here, for simplicity, we consider $\Gamma _{bs} ^R=0$ and the non zero $Z^{\prime} -b-s$ coupling ($\Gamma_{bs}^L$) is taken to be real in our analysis. 

\subsection{Constraints on $Z^{\prime}$ couplings from leptonic decays}
In lepton flavor violating $\tau$ decays, the $\tau \to \ell \ell \ell$ ($\ell = \mu, e$) channel provides very sensitive probe of the coupling $\Gamma _{\ell \ell ^{\prime}}$ in $Z^{\prime}$ model. Experimentally, the branching ratio for the process $\tau \rightarrow \mu\mu\mu$ and $\tau \rightarrow eee$ processes are $2.1\times 10^{-8}$ and $2.7\times 10^{-8}$ at $90\%$ CL, respectively~\cite{ParticleDataGroup:2022pth}. Additionally, an upper bound of the branching fraction of $\mu \to eee$ process is $1.0\times 10^{-12}$\cite{ParticleDataGroup:2022pth}.  However, a significant sensitivity in beyond the SM could be provided by additional observation of such distinct processes in the collider experiments.
The $Z^{\prime}$ boson contributes to the LFV 3-body leptonic decay $\tau \rightarrow \mu\mu\mu$ at the tree level with the branching ratio which is given by~\cite{Langacker:2008yv, Crivellin:2013hpa}
\bea
\mathcal{B}(\tau \rightarrow \mu\mu\mu) =\frac{m_{\tau}}{1536\pi^3\Gamma_{\tau}M_{Z'}^4}
[2(|\Gamma_{\mu\tau}^L \Gamma_{\mu\mu}^L|^2 + |\Gamma_{\mu\tau}^R \Gamma_{\mu\mu}^R|^2) 
+ |\Gamma_{\mu\tau}^L\Gamma_{\mu\mu}^R|^2 
+ |\Gamma_{\mu\tau}^R\Gamma_{\mu\mu}^L|^2],
\eea
where the $m_\tau~(\Gamma _\tau)$ is the mass (total decay width) of $\tau$ lepton. 
\begin{table}[h]
\begin{center}
\begin{tabular}{|c | c | c| c |}
\hline
 Couplings  & $m_{Z^{\prime}}=4.5$ TeV &  $m_{Z^{\prime}}=6.0$ TeV &   $m_{Z^{\prime}}=7.0$ TeV\\
 \hline
 \hline
 $\Gamma _{\tau \mu}(\mathcal{S} - \rm I)$  & 0.128  & 0.227 & 0.310 \\
  $\Gamma _{\tau e}(\mathcal{S} - \rm I)$ & 0.145 & 0.258 & 0.351 \\
 $\Gamma _{\mu e}(\mathcal{S} - \rm I)$ & $3.230 \times 10^{-4}$  & $5.742 \times 10^{-4}$ & $7.815 \times 10^{-4}$ \\
\hline
\hline
  $\Gamma _{\tau \mu}(\mathcal{S} - \rm II)$ & 0.221  & 0.394 & 0.537\\
$\Gamma _{\tau e}(\mathcal{S} - \rm II)$ & 0.251 & 0.447 & 0.609	 \\
$\Gamma _{\mu e}(\mathcal{S} - \rm II)$ & $4.568 \times 10^{-4}$  & $8.120 \times 10^{-4}$ & $1.105 \times 10^{-3}$  \\

\hline
\end{tabular}
\end{center}
\caption{The NP couplings for LFV leptonic decays }
\end{table} {\label{LFVLC3}}
Similarly, the branching ratio of $\tau \to eee$  decay mode can be obtained by replacing  $\mu$ by $e$.

The LFV branching fraction of $\mu \to eee$ decay mode is given as~\cite{Dib:2018rpy} 

\bea
\mathcal{B}(\mu \to eee) = \frac{g_{V\mu e}^2g_{Ve e}^2}{m_{Z^{\prime}}^4}\frac{2}{4G_F^2},
\eea
where $g_V^2= |g_{\mu e}^V|^2+|g_{\mu e}^A|^2$,  and $G_F$ is the Fermi coupling constant.

Using the upper limits on the above discussed LFV leptonic decays, we obtain the values of lepton flavor violating NP couplings as given in Table - (\ref{LFVLC3}) where the coupling of $Z^{\prime}$ to $\ell \ell$ is considered as SM like in this analysis.


\subsection{Fit Results}

In this sub-section, we consider the $q^2$ bin SM and experimental measurements of various observables that include the angular observable $P_5'$,  $ \mathcal{B} (B_s \to \phi \mu \mu)$ and $\mathcal{B} (B_s \to \mu \mu)$ . The form factor independent observable $P_5'$ in $B\to K^* \mu \mu$ process is defined as
\bea
P_5' = \frac{J_5}{2\sqrt{-J_2^c J_2^s}},
\eea
where the auxiliary functions $J_i^p (i =2,5; p = c, s)$ includes the relevant form factors of the transition $B \to K^*$ and the Wilson coefficients. For the numerical calculation of $B\to K^* \ell \ell$, we employ the FF from the LQCD method~\cite{Bouchard:2013eph}. Similarly, $B_s \to \phi \mu \mu$ processes induced by FCNC $b\to s \mu \mu$ transition, we consider the FFs from the combined analysis of the LCSR  and LQCD fit results~\cite{Bharucha:2015bzk}. The branching ratio of $ B_s \to \mu \mu$ leptonic decay mode has also been studied. Now, using these three observables, we perform a naive $\chi^2$ analysis to obtain the NP coupling parameters for the non-universal $Z'$ model. We define the $\chi ^2$ formula which is defined as follows
\bea
\chi^2(C_i^{\rm NP)}= \sum_i  \frac{\Big ({\cal O}_i^{\rm th}(C_{9,10}^{\rm NP}) -{\cal O}_i^{\rm exp} \Big )^2}{(\Delta {\cal O}_i^{\rm exp})^2+(\Delta {\cal O}_i^{\rm sm})^2},
\eea
where, ${\cal O}_i ^ {\rm th}$ represents the theoretical expressions including the NP contributions and ${\cal O}_i ^ {\rm exp}$ are the experimental values. The denominator  includes $1 \sigma$ uncertainties associated with the theoretical and experimental results. As the new vector boson $Z^{\prime}$ is not yet observed in collider experiments, its mass scale is constrained in different Grand unified theories and discussed in Refs.~\cite{ATLAS:2017eiz, CMS:2018ipm, Bandyopadhyay:2018cwu}. Bandopadhyay et al. have constrained as $m_{Z^{\prime}}>4.4$ TeV using the recent Drell-Yan data of the LHC~\cite{Bandyopadhyay:2018cwu}.
By using all the input parameters, the values of the NP parameters are given below
\bea
 \Gamma_{bs}^ L|_{\mathcal{S}- \rm I} &=&0.060 ~(m_{Z^{\prime}}=4.5~ \rm TeV), 0.108 ~(m_{Z^{\prime}}=6.0~ \rm TeV), 0.147 ~(m_{Z^{\prime}}=7.0~ \rm TeV),\nn\\
 \Gamma_{bs}^ L|_{\mathcal{S}- \rm II} &=&0.062 ~(m_{Z^{\prime}}=4.5~ \rm TeV),0.110 ~(m_{Z^{\prime}}=6.0~ \rm TeV),  0.150
 ~(m_{Z^{\prime}}=7.0~ \rm TeV),
\eea
where we have used three $m_{Z^{\prime}}$ values in this analysis.
\subsection{Input parameters}
In this sub-section, we employ all the input parameters used for our analysis. From Ref.~\cite{ParticleDataGroup:2020ssz}, we consider all the necessary parameters such as CKM matrix element, life time of $B_{(s)}$ meson, Fermi coupling constant, fine structure constant, and masses of quarks, and leptons, etc. We employ the Wilson coefficients at the scale $\mu =m_b$  from Ref.~\cite{Ali:1999mm}. 

\begin{table}[htp]
\centering
\scalebox{0.9}{
\begin{tabular}{|c|c|c|c|c|c|c|}
\hline
Observable & & $m_{Z^{\prime}}$=4.5 TeV & $m_{Z^{\prime}}$=6.0 TeV & $m_{Z^{\prime}}$=7.0 TeV \\
\hline
\hline
\multicolumn{5}{|c|}{$B_s \to \ell \ell ^{\prime}$ ($Z'$ contribution)}\\
\hline
\hline
\multirow{2}{*}{$\mathcal{B}_ {\mu e}\times 10^{-9}$} 
& $\mathcal{S} - \rm I$ & $4.041 \times 10^{-9}$ & $1.271 \times 10^{-8}$ & $2.354 \times 10^{-8}$  \\
\cline{2-5}
& $\mathcal{S} - \rm II$ & $1.632 \times 10^{-8}$ & $5.083 \times 10^{-8}$ & $7.248 \times 10^{-8}$  \\

\hline $\mathcal{B}_ {\tau \mu}\times 10^{-8}$ 
& $\mathcal{S} - \rm I$ & $0.068$ & $0.240$ & $0.449$   \\
\cline{2-5}
& $\mathcal{S} - \rm II$ & $0.485$ & $1.623$ & $3.026$   \\
\hline
\multirow{2}{*}{$\mathcal{B}_ {\tau e}\times 10^{-9}$} 
& $\mathcal{S} - \rm I$ & $0.196$ & $0.617$ & $1.143$  \\
\cline{2-5}
& $\mathcal{S} - \rm II$ & 0.406 & 1.282 & 1.832  \\
\hline
\hline
\end{tabular}
}
\caption{Upper limit values of $ B_s \to \ell \ell ^{\prime}$ processes}
\label{tab_Btollp}
\end{table}
\section{Numerical analysis and discussions}\label{NAD}
\subsection{Analysis of $B_s \to \ell \ell^{\prime}$ process}
We estimate the numerical values of the branching ratios of $B_s \to \ell \ell ^{\prime}$ processes and shown in Table.~\ref{tab_Btollp}. One can observe that the branching fraction of $B_s \to \mu e$ mode is highly suppressed as compared to $\tau \mu$ and $\tau e$ channels ($\mathcal{O}(10^{-9})$) present in the final state.  We show for three $m_{Z^{\prime}}$ values for the given processes and obtain that the contribution increases in the presence of the NP couplings.
\begin{table}[htp]
\centering
\scalebox{0.9}{
\begin{tabular}{|c|c|c|c|c|c|c|}
\hline
Observable & & $m_{Z^{\prime}}$=4.5 TeV & $m_{Z^{\prime}}$=6.0 TeV & $m_{Z^{\prime}}$=7.0 TeV \\
\hline
\hline
\multicolumn{5}{|c|}{$B_s \to K \ell \ell ^{\prime}$ ($Z'$ contribution)}\\
\hline
\hline
\multirow{2}{*}{$\mathcal{B}_ {\mu e}\times 10^{-13}$} 
& $\mathcal{S} - \rm I$ & $0.221$ & $0.697$ & $1.291$  \\
\cline{2-5}
& $\mathcal{S} - \rm II$ & 0.919 & 2.892 & 5.376  \\

\hline $\mathcal{B}_ {\tau \mu}\times 10^{-8}$ 
& $\mathcal{S} - \rm I$ & $0.221$ & $0.693$ & $1.292$   \\
\cline{2-5}
& $\mathcal{S} - \rm II$ & $0.354$ & $1.126$ & $2.096$   \\
\hline
\multirow{2}{*}{$\mathcal{B}_ {\tau e}\times 10^{-8}$} 
& $\mathcal{S} - \rm I$ & $0.292$ & $0.921$ & $1.705$  \\
\cline{2-5}
& $\mathcal{S} - \rm II$ & 0.458 & 1.453 & 2.702 \\
\hline
\hline
\end{tabular}
}
\caption{Estimated upper limit values of $ B \to K \ell \ell ^{\prime}$ processes in $Z^{\prime}$ model} 
\label{tab_BtoK}
\end{table}
\begin{figure}[htb]
\includegraphics[scale=0.57]{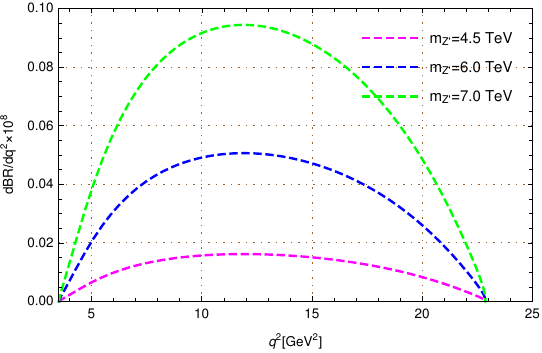} 
\quad
\includegraphics[scale=0.57]{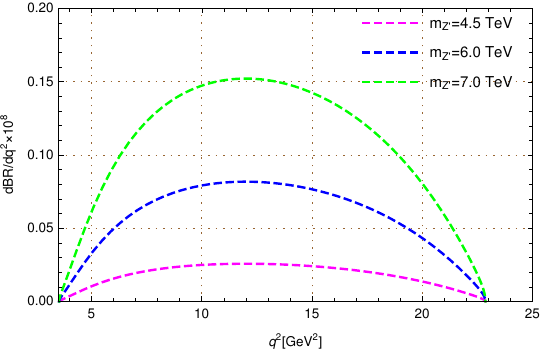} 
\quad
\includegraphics[scale=0.57]{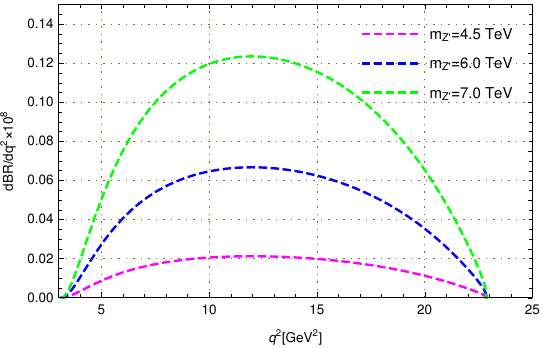} 
\quad
\includegraphics[scale=0.57]{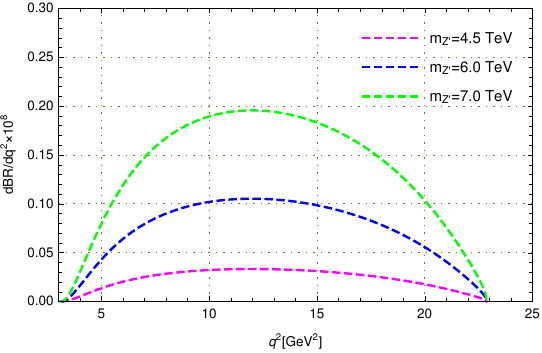}
\quad 
\includegraphics[scale=0.57]{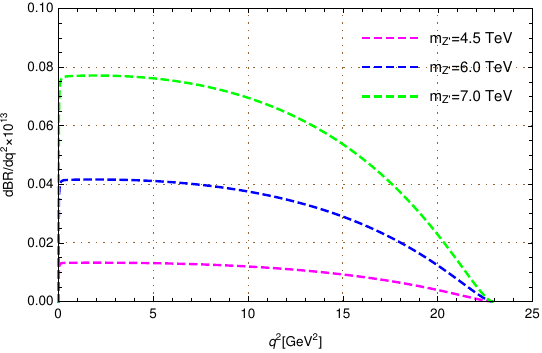} 
\quad
\includegraphics[scale=0.57]{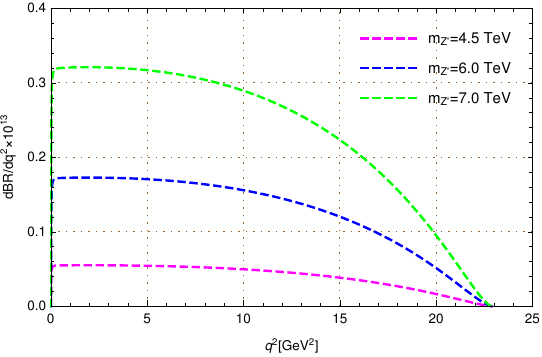} 
\caption{Variation of branching ratios of $B\to K \tau \mu$ (top), $B\to K \tau e$ (middle), $B \to K \mu e$ (bottom) processes in $Z^{\prime}$ model. The left (right) panel corresponds to $\mathcal{S}- {\rm I}$ (${\rm II}$).} \label{BtoK}
\end{figure}
\subsection{Analysis of $B \to K \ell \ell^{\prime}$ process}
After having knowledge about the NP coupling in details, we now proceed to analyse the above discussed prominent observables of $B \to K \ell \ell ^{\prime}$ lepton flavor violating process mediated by $b\to s \ell \ell^{\prime}$ transition in the $Z^{\prime}$ model. 

In Fig. \ref{BtoK}, we show the variation of the differential branching ratio of $B \to K \tau \mu$ (top-left), $B \to K \tau e$ (middle-left) and $B \to K \mu e$ (bottom) processes w.r.t $q^2$ in the presence of $Z ^{\prime}$ model. Here the magenta, blue, and green lines represent the contributions in the $Z^{\prime}$ model with three different values of $m_{Z^{\prime}}$. We observe that the observables have a higher contribution in the mid $q^2$ region for $B\to K\tau \mu$ and $B\to \tau e$ processes, and in low $q^2$ region for $B\to K\mu e$ transition. This behavior arises due to the lighter lepton masses involved in the later mode. However, the presence of NP does help to enhance the contribution while increasing the $m_{Z^{\prime}}$ values. The predicted branching fractions are shown in Table.~\ref{tab_BtoK}. This indicates that the branching fraction of $B \to K \mu e$ channel is very suppressed as compared to the $\tau \mu$ and $\tau e$ final states. 
\begin{figure}[htb]
\includegraphics[scale=0.57]{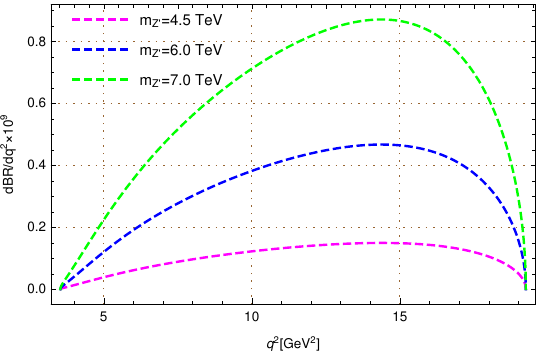} 
\quad
\includegraphics[scale=0.57]{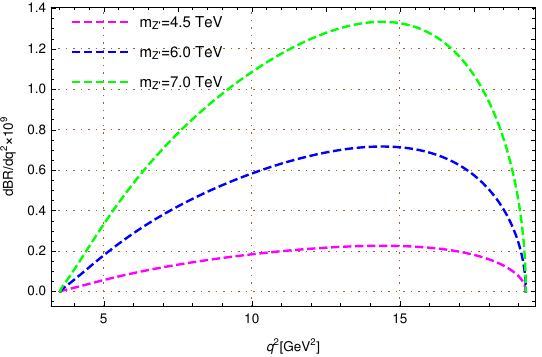} 
\quad
\includegraphics[scale=0.57]{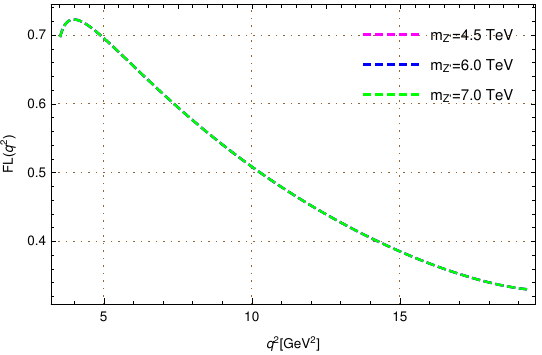}
\quad
 \includegraphics[scale=0.57]{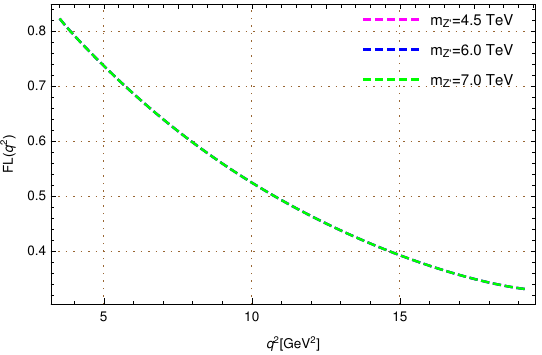}
\quad
\includegraphics[scale=0.57]{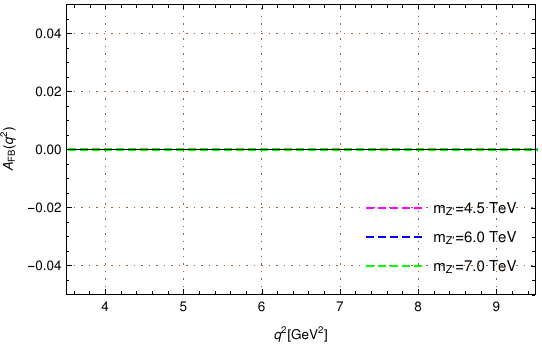} 
\quad
\includegraphics[scale=0.57]{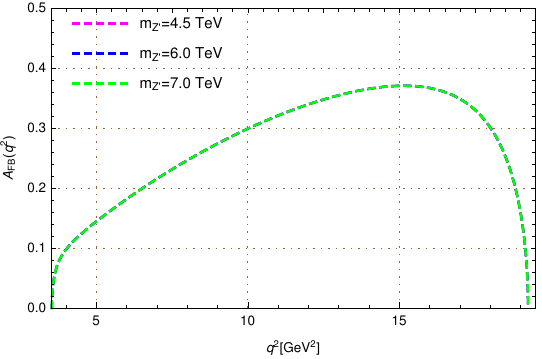} 
\caption{$B\to K^*\tau \mu$: $d\mathcal{B}/dq^2$,$F_L (q^2)$ and $A_{FB} (q^2)$ for $\mathcal{S}- {\rm I}$ (left panel) and $\mathcal{S}- {\rm II}$ (right panel).} {\label{BtoKstaumu}}
\end{figure}
 In this scenario, we display the  $q^2$ behavior of the branching ratios of $B\to K \tau \mu$  and $B\to K \tau e$ processes in the top-right and middle-right panels, respectively shown in Fig.~\ref{BtoK}. In the observable, we consider all the central values of the input parameters and form factors. The colour description of the figures is same as the $\mathcal{S} - {\rm I}$. In the variation of whole $q^2$ region, the contributions in the presence of NP coupling indicate that the observable has higher values in the mid-region. However, the presence of the NP coupling increase the central values with same order $\mathcal{O}(10^{-8})$. Table - \ref{tab_BtoK} summarizes the estimated branching fractions in the whole kinematic region.
\begin{table}[htp]
\centering
\scalebox{0.7}{
\begin{tabular}{|c|c|c|c|c|c|c|c|c|}
\hline
Observable & & $m_{Z^{\prime}}$=4.5 TeV & $m_{Z^{\prime}}$= 6.0 TeV & $m_{Z^{\prime}}$=7.0 TeV & $m_{Z^{\prime}}$=4.5 TeV & $m_{Z^{\prime}}$= 6.0 TeV & $m_{Z^{\prime}}$=7.0 TeV \\
\hline
\hline
\multicolumn{2}{|c|} {\textcolor{white}{.}} & \multicolumn{3}{|c|}{$B_s \to K^* \ell \ell ^{\prime}$ ($Z'$ contribution)} & \multicolumn{3}{|c|}{$B_s \to \phi \ell \ell ^{\prime}$ ($Z'$ contribution)}\\
\hline
\multirow{2}{*}{$\mathcal{B}_ {\mu e}\times 10^{-13}$} 
& $\mathcal{S} - \rm I$ & $0.193$ & $0.608$ & $1.126$ & 0.523 & 1.647 & 3.052  \\
\cline{2-8}
& $\mathcal{S} - \rm II$ & 0.801 & 2.523 & 4.691 & 2.095 &  6.589& 12.203\\
\hline
\multirow{2}{*}{$\mathcal{B}_ {\tau \mu}\times 10^{-9}$} 
& $\mathcal{S} - \rm I$ & $1.654$ & $5.178$ & $9.658$ & 4.107 & 12.856 & 23.976 \\
\cline{2-8}
& $\mathcal{S} - \rm II$ & $2.492$ & $7.912$ & $14.723$  & 6.185 & 19.638 & 36.541 \\
\hline
\multirow{2}{*}{$\mathcal{B}_ {\tau e}\times 10^{-9}$} 
& $\mathcal{S} - \rm I$ & $2.058$ & $6.483$ & $12.000$ & 5.111 & 16.103 & 29.805 \\
\cline{2-8}
& $\mathcal{S} - \rm II$ & $3.226$ & $10.222$ & $19.006$ & 8.013 & 25.388 & 47.203 \\
\hline
\multirow{2}{*}{$\mathcal{F}_L^{\mu e}$} 
& $\mathcal{S} - \rm I$ & $0.488$ & $0.488$ & $0.488$ & 0.552 & 0.552 & 0.552 \\
\cline{2-8}
& $\mathcal{S} - \rm II$ & 0.488 & 0.488 & 0.488 & 0.552 & 0.552 & 0.552  \\
\hline
\multirow{2}{*}{$\mathcal{F}_L^{\tau \mu}$} 
& $\mathcal{S} - \rm I$ & 0.456 & 0.456 & 0.456  & 0.501 & 0.501 & 0.501 \\
\cline{2-8}
& $\mathcal{S} - \rm II$ & 0.468 & 0.468 & 0.468 & 0.514 & 0.514 & 0.514\\
\hline
\multirow{2}{*}{$\mathcal{F}_L^{\tau e}$} 
& $\mathcal{S} - \rm I$ & 0.468 & 0.468 & 0.468 & 0.514 & 0.514 & 0.514 \\
\cline{2-8}
& $\mathcal{S} - \rm II$ & 0.468 & 0.468 & 0.468  & 0.514 & 0.514 & 0.514 \\
\hline
\multirow{2}{*}{$\mathcal{A}_{FB}^{\mu e}$} 
& $\mathcal{S} - \rm I$ & 0.000 & 0.000 & 0.000 & 0.000 & 0.000 & 0.000\\
\cline{2-8}
& $\mathcal{S} - \rm II$ & 0.349 & 0.349 & 0.349  & 0.295 & 0.295 & 0.295\\
\hline
\multirow{2}{*}{$\mathcal{A}_{FB}^{\tau \mu}$} 
& $\mathcal{S} - \rm I$ & 0.000 & 0.000 & 0.000  & 0.000 & 0.000 & 0.000 \\
\cline{2-8}
& $\mathcal{S} - \rm II$ & 0.311 & 0.311 & 0.311 & 0.271 &  0.271 &  0.271 \\
\hline
\multirow{2}{*}{$\mathcal{A}_{FB}^{\tau e}$} 
& $\mathcal{S} - \rm I$ & 0.000 & 0.000 & 0.000 & 0.000 & 0.000 &  0.000 \\
\cline{2-8}
& $\mathcal{S} - \rm II$ & 0.311 & 0.311 & 0.311 & 0.272 & 0.272 &  0.272 \\
\hline
\hline
\end{tabular}
}
\caption{Upper limit values of $ B \to (K^*, \phi) \ell \ell ^{\prime}$ processes in $Z^{\prime}$ model}
\label{tab_BtoKsphi}
\end{table}
\subsection{Analysis of $B \to V(K^*,\phi) \ell \ell^{\prime}$ process}
In this sub-section, we provide a detailed study of $B \to V \ell \ell^{\prime}$ processes mediated by $b \to s \ell \ell^{\prime}$ quark level transition where the vector meson $V=K^*,~\phi$. We probe the NP effects on the associated observables such as differential branching ratio ($d\mathcal{B}/dq^2$), the forward-backward asymmetry ($A_{FB}$), and the longitudinal polarisation fraction ($F_L$). 

\begin{figure}[htb]
\includegraphics[scale=0.57]{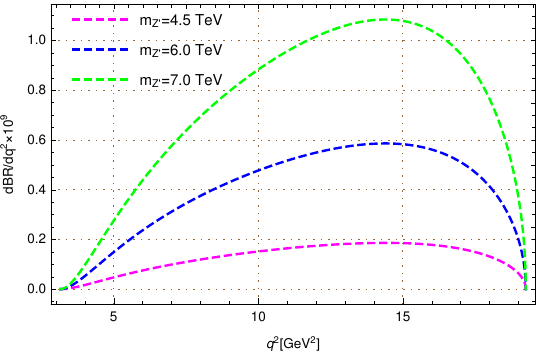} 
\quad
\includegraphics[scale=0.57]{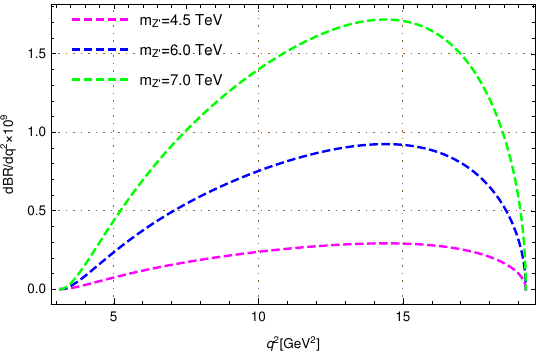} 
\quad
\includegraphics[scale=0.57]{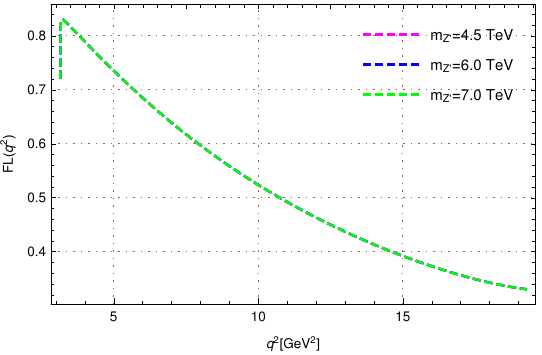}
\quad
 \includegraphics[scale=0.57]{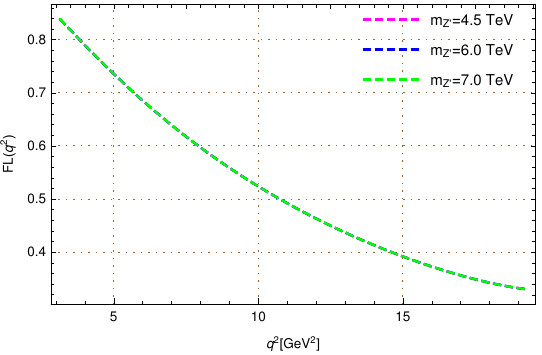}
\quad
\includegraphics[scale=0.57]{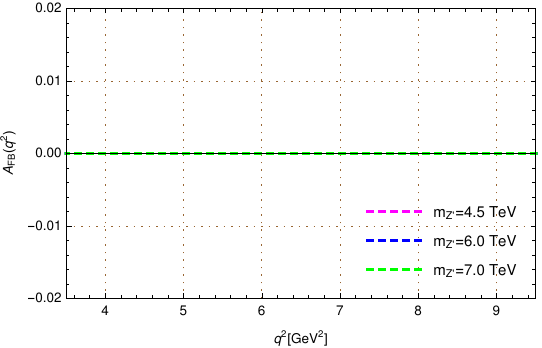} 
\quad
\includegraphics[scale=0.57]{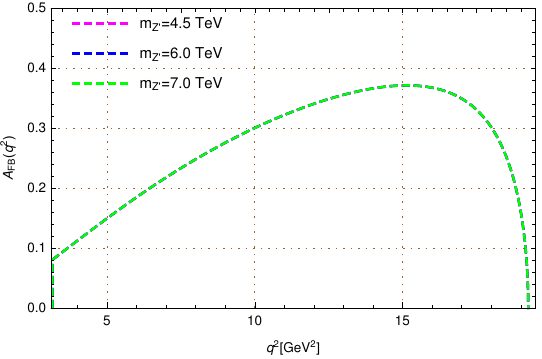} 
\caption{$B\to K^*\tau e$: $d\mathcal{B}/dq^2$,$F_L (q^2)$, $A_{FB} (q^2)$ for $\mathcal{S}- {\rm I}$ (left panel) and $\mathcal{S}- {\rm II}$ (right panel).} {\label{BtoKstaue}}
\end{figure}
\begin{figure}[htb]
\centering
\includegraphics[scale=0.57]{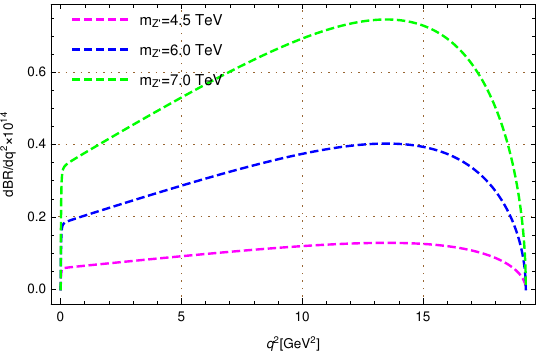}
\quad
\includegraphics[scale=0.57]{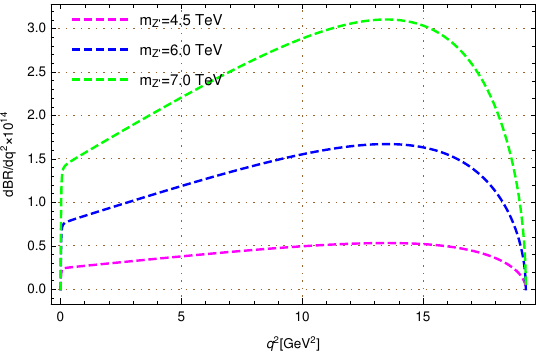}
\quad
\includegraphics[scale=0.57]{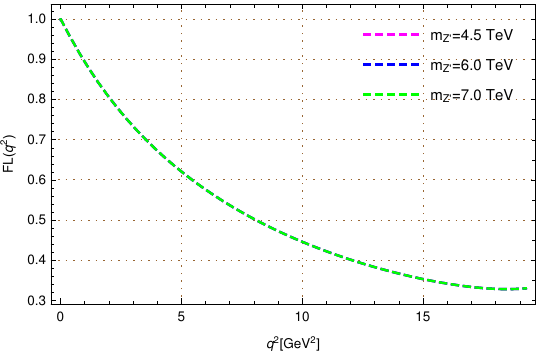} 
\quad
\includegraphics[scale=0.57]{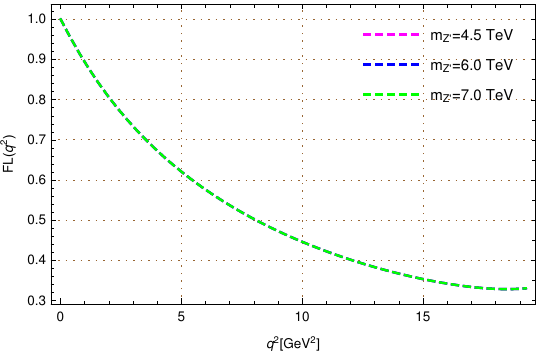} 
\quad
\includegraphics[scale=0.57]{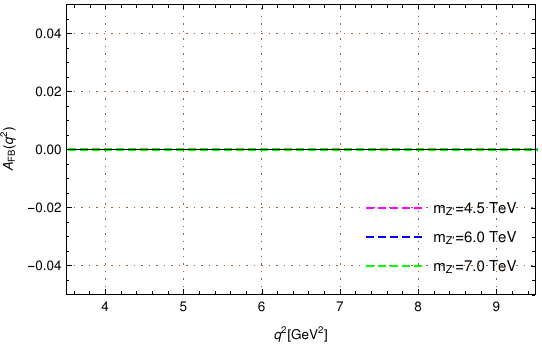} 
\quad
\includegraphics[scale=0.57]{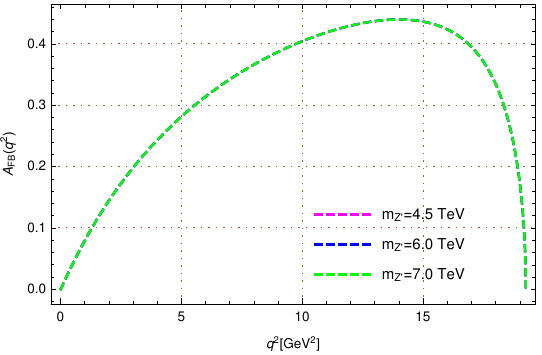} 
\caption{ The observables $d\mathcal{B}/dq^2$, $F_L (q^2)$ and $A_{FB} (q^2)$ of $B\to K^*\mu e$ process in scenario -{$\rm I$}(left panel) and scenario -{$\rm II$} (right panel).} {\label{BtoKsmue}}
\end{figure}
\begin{figure}[htb]
\includegraphics[scale=0.561]{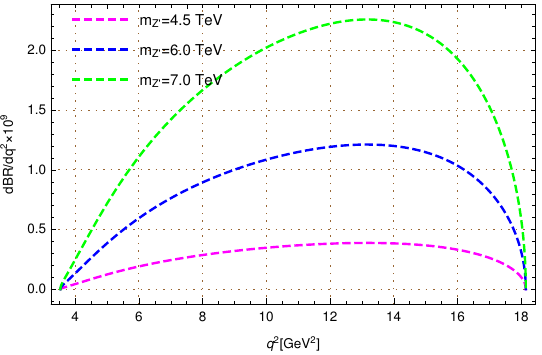} 
\quad
\includegraphics[scale=0.561]{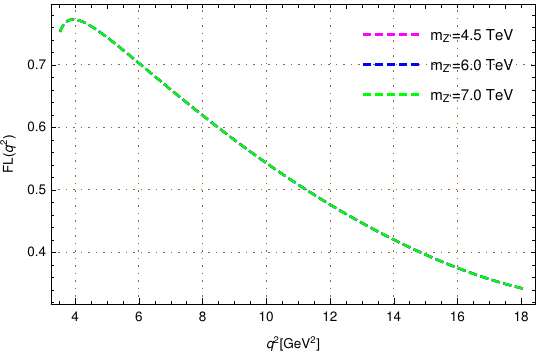} 
\quad
\includegraphics[scale=0.561]{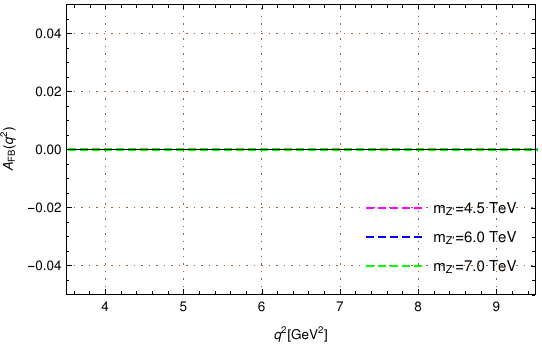}
\quad
 \includegraphics[scale=0.561]{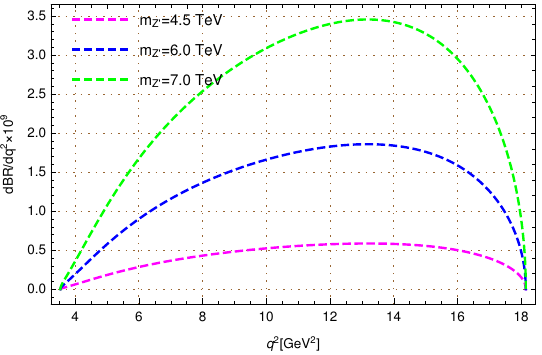}
\quad
\includegraphics[scale=0.561]{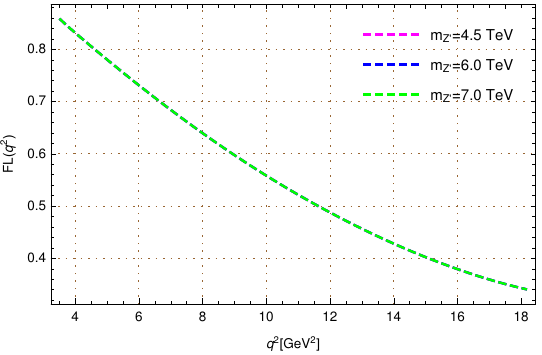} 
\quad
\includegraphics[scale=0.561]{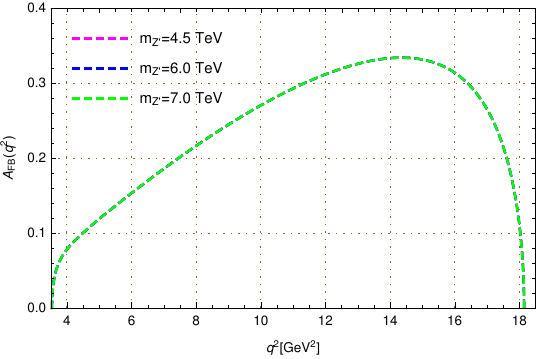} 
\caption{Variation of BR (top-left),$F_L$ (top-middle), $A_{FB}$ (top-right) of $B\to \phi\tau \mu$ process in $\mathcal{S}- {\rm I}$ and the bottom panel (left, middle and right) depicts for $\mathcal{S}- {\rm II}$.}{\label{Btophitaumu}}
\end{figure}
\begin{figure}[htb]
\includegraphics[scale=0.561]{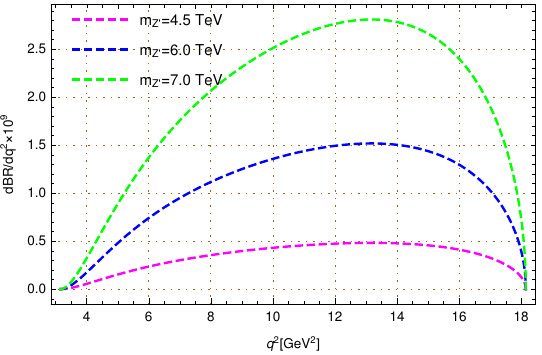} 
\quad
\includegraphics[scale=0.561]{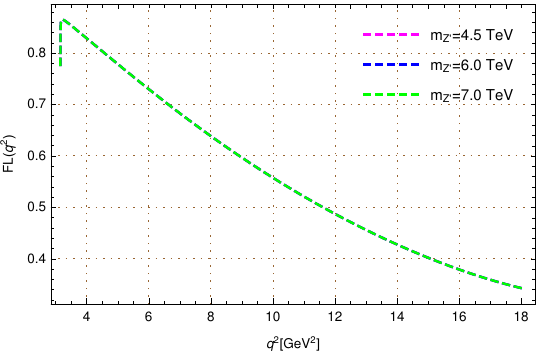} 
\quad
\includegraphics[scale=0.561]{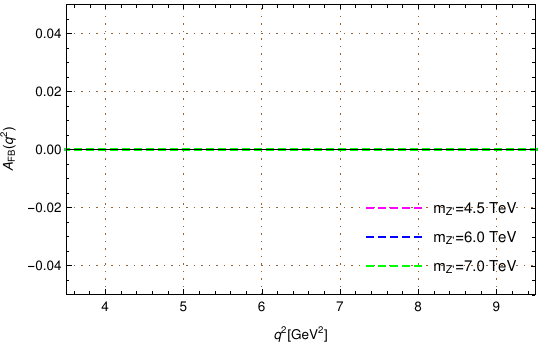}
\quad
 \includegraphics[scale=0.561]{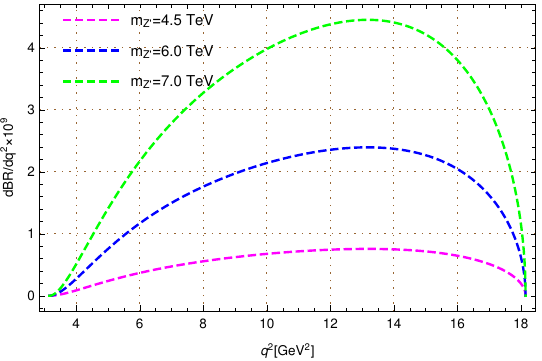}
\quad
\includegraphics[scale=0.561]{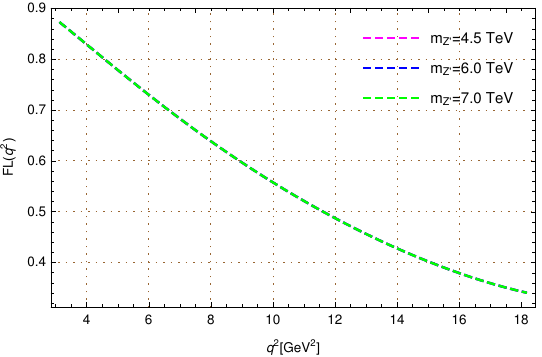} 
\quad
\includegraphics[scale=0.561]{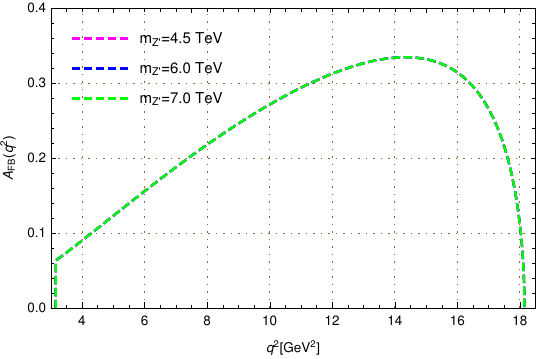} 
\caption{The $q^2$ variationn of BR (top-left),$F_L$ (top-middle), $A_{FB}$ (top-right) of $B\to \phi\tau e$ process in $\mathcal{S}- {\rm I}$, and the bottom panel (left, middle and right) depicts for $\mathcal{S}- {\rm II}$.}{\label{Btophitaue}}
\end{figure}
\begin{figure}[htb] 
\centering
\includegraphics[scale=0.561]{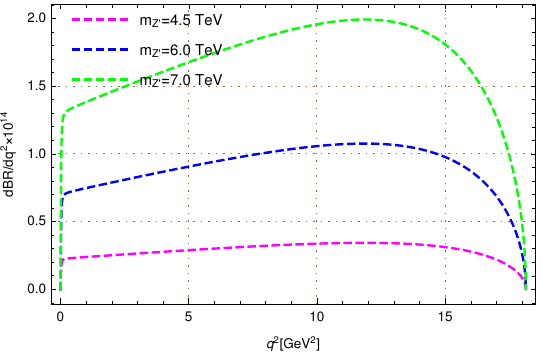}
\quad
\includegraphics[scale=0.561]{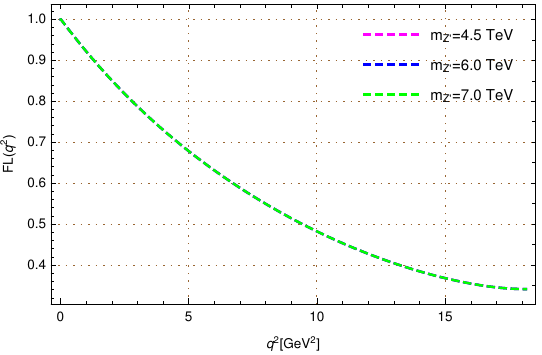}
\quad
\includegraphics[scale=0.561]{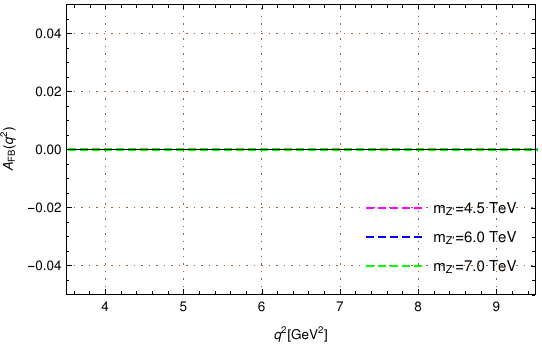} 
\quad
\includegraphics[scale=0.561]{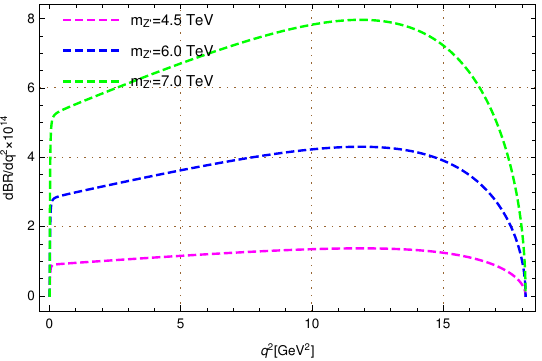} 
\quad
\includegraphics[scale=0.561]{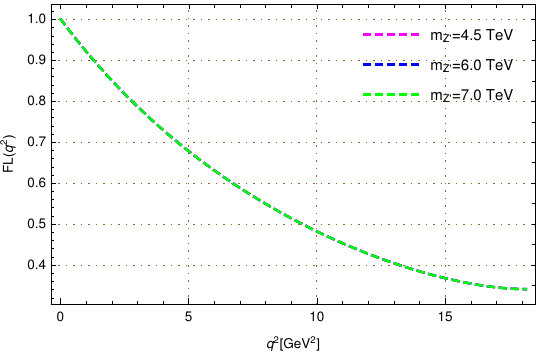} 
\quad
\includegraphics[scale=0.561]{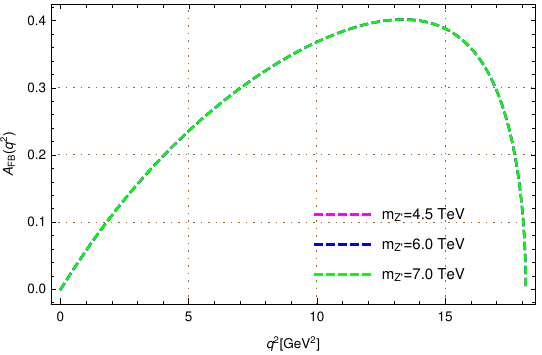} 
\caption{The $q^2$ variation of BR (top-left),$F_L$ (top-middle), $A_{FB}$ (top-right) of $B\to \phi\mu e$ process in $\mathcal{S}- {\rm I}$ and the bottom panel (left, middle and right) depicts for $\mathcal{S}- {\rm II}$.}{\label{Btophimue}}
\end{figure}

In the right panels of Fig.~\ref{BtoKstaumu} and \ref{BtoKstaue}, we analyse the $q^2$ variation of the above discussed observables of $B\to K^* \tau \mu$ and $B\to K^* \tau e$ processes with respect to $q^2$, respectively. The color description of the plots are same as previous. The $q^2$ dependent branching ratios $d\mathcal{B}/dq^2$ have distinguished contributions in the presence of NP couplings. Higher the values of $m_{Z^{\prime}}$ induce the larger contributions to the observable. In the middle - (left, right) panel, however, in the variation of the sensitive observable $FL(q^2)$, the contribution in presence of NP couplings are indistinguished and coincides for all $m_{Z^{\prime}}$ entries. In the observable $A_{FB}(q^2)$ shown in bottom - left panel, the NP contribution allows no contribution to the NP in scenario - $\rm I$ whereas a definite contribution arises from scenario - $\rm II$. We also show the $q^2$ variation of the branching ratio of $B \to K^* \mu e$ process, shown in Fig.~\ref{BtoKsmue}. 
In the low $q^2$ region, the NP contribution in the presence of $Z^{\prime}$ couplings gets higher and varies remarkably w.r.t $q^2$ for three values of $m_{Z^{\prime}}$. In the study of whole kinematic region of $q^2$, the lepton polarisation asymmetry observable decreases but doesn't drop at zero point. On the other hand, the observable $A_{FB}(q^2)$ varies in the whole $q^2$ kinematic region in scenario - ${\rm I}$ whereas it provides a significant contribution in scenario - $\rm II$. The associated plots have been depicted in Fig.~\ref{BtoKsmue}. 
We also present the values of the branching ratios in the whole kinematic region as shown in Table.~\ref{tab_BtoKsphi}. The branching ratios of $B \to K^* \tau \mu$ and $B \to K^* \tau e$ processes are of the same order i.e $\mathcal{O} (10^{-9})$ and differ only in their central values whereas the branching fraction of $B \to K^* e \mu$ process are suppressed with $\mathcal{O} (10^{-13})$. The forward-backward asymmetry, and the polarisation asymmetry observables differ in all the above discussed processes individually.
Similar to the $B\to K^* \ell \ell^{\prime}$ process we investigate another decay channel  $B_s \to \phi \ell \ell ^{\prime}$. We depict the $q^2$ dependent physical observables such as branching ratio, forward-backward asymmetry, and polarisation asymmetry for $B_s \to \phi \tau \mu$ and $B_s \to \phi \tau e$ in Fig.~\ref{Btophitaumu} and \ref{Btophitaue}, respectively whereas Fig.~\ref{Btophimue} indicates the variation of the above discussed observables of $B_s \to \phi \mu e$ decay channel. The top-left, top-middle, and top-right panels are shown for branching ratio, polarisation asymmetry and forward-backward asymmetry for scenario - $\rm I$, respectively. Similarly, the bottom-left, bottom-middle, and bottom-right panels depict for the given observables in scenario - $\rm II$.
In the presence of NP couplings from $Z^{\prime}$ model, we obtain similar results compared to previous channel with difference in the variation due to the masses of mesons and the transition form factor involved in this analysis.
\newpage
\newpage
\subsection{Analysis of $B \to T(K_2^*,f_2^{\prime}) \ell \ell^{\prime}$ process}
Here, we study the exclusive semileptonic lepton flavor violating $B \to T(K_2^*,f_2^{\prime}) \ell \ell^{\prime}$ channels in details in the framework of non-universal $Z^{\prime}$ model. Similar to the previous processes, we also analyse in the scanario -$\rm I$ and $\rm II$.
\begin{figure}[htb]
\centering
\includegraphics[scale=0.57]{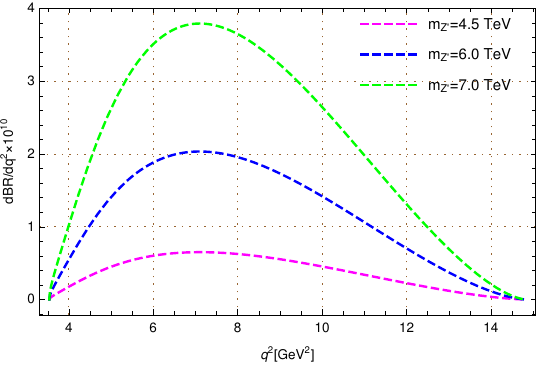}
\quad
\includegraphics[scale=0.57]{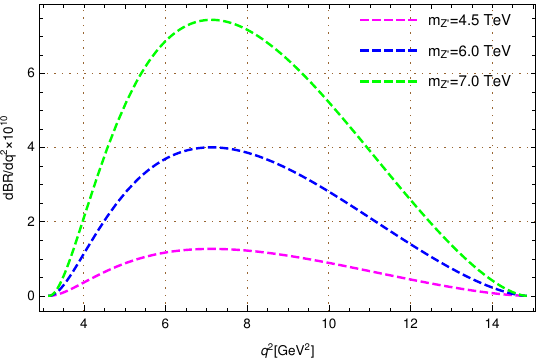}
\quad
\includegraphics[scale=0.57]{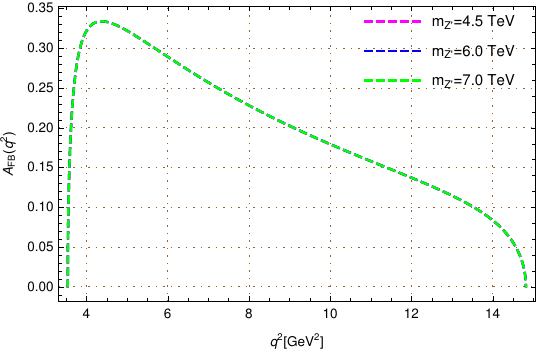} 
\quad
\includegraphics[scale=0.57]{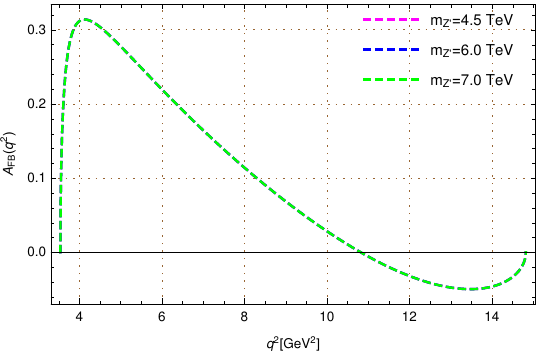}
\caption{The variation of $\mathcal{B}$ and $ A_{FB}$ of $B\to K_2^* \tau \mu$ process are shown in $\mathcal{S}- {\rm I}$ (left panel) and $\mathcal{S}- {\rm II}$ (right panel).} {\label{BtoK2staumu}}
\end{figure}
\begin{figure}[htb]
\centering
\includegraphics[scale=0.57]{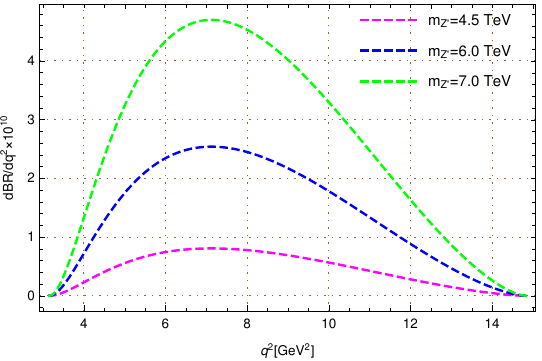}
\quad
\includegraphics[scale=0.57]{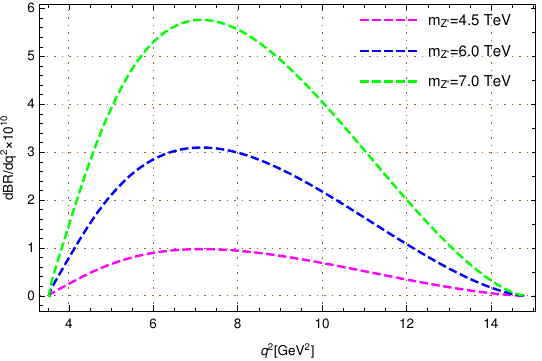}
\quad
\includegraphics[scale=0.57]{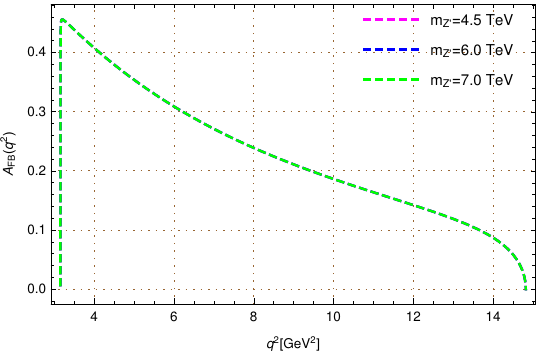} 
\quad
\includegraphics[scale=0.57]{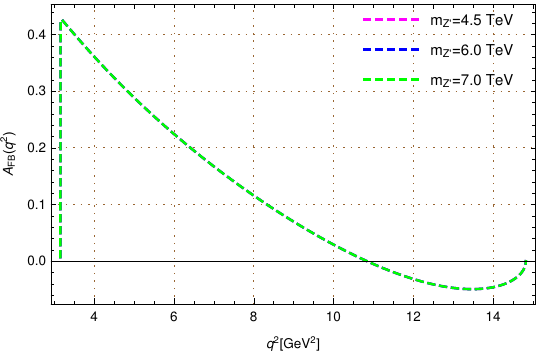}
\caption{The variation of the $\mathcal{B}$ and $ A_{FB}$ of $B\to K_2^* \tau e$ process in the $Z^{\prime}$ model: Left and right panels are for scenario - ${\rm I}$ and ${\rm II}$, respectively.} {\label{BtoK2staue}}
\end{figure}
\begin{figure}[htb]
\centering
\includegraphics[scale=0.57]{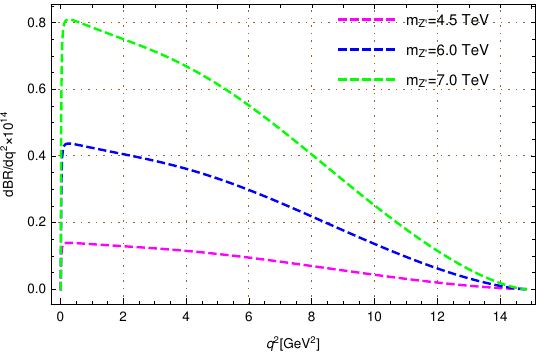}
\quad
\includegraphics[scale=0.57]{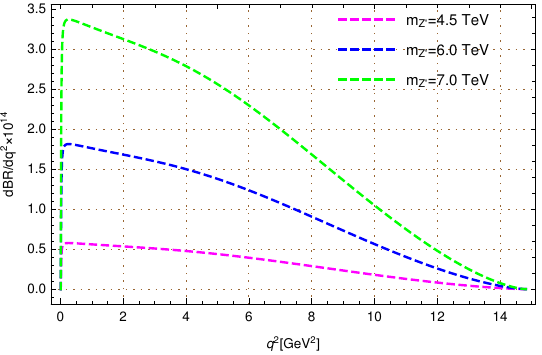}
\quad
\includegraphics[scale=0.57]{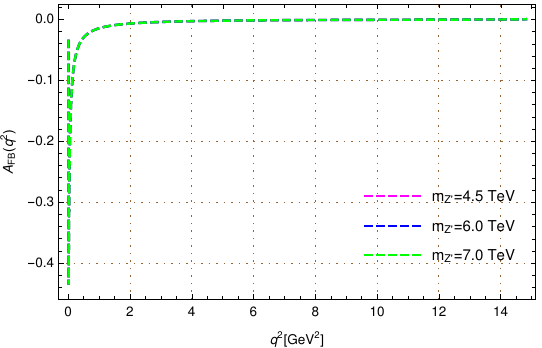}
\quad
\includegraphics[scale=0.57]{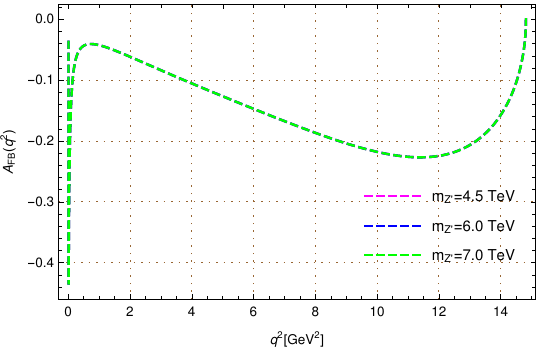}
\caption{The variation of differential branching ratio and forward-backward asymmetry of $B\to K_2^* \mu e$ process with respect to $q^2$. The left (right) panel indicates $\mathcal{S} - {\rm I} ({\rm II})$.}{\label{BtoK2smue}}
\end{figure}
\begin{table}[htp]
\centering
\scalebox{0.7}{
\begin{tabular}{|c|c|c|c|c|c|c|c|c|}
\hline
Observable & & $m_{Z^{\prime}}$=4.5 TeV & $m_{Z^{\prime}}$=6.0 TeV & $m_{Z^{\prime}}$=7.0 TeV & $m_{Z^{\prime}}$=4.5 TeV & $m_{Z^{\prime}}$=6.0 TeV & $m_{Z^{\prime}}$=7.0 TeV \\
\hline
\hline
\multicolumn{2}{|c|} {\textcolor{white}{.}} & \multicolumn{3}{|c|}{$B_s \to K_2^* \ell \ell ^{\prime}$ ($Z'$ contribution)} & \multicolumn{3}{|c|}{$B \to f_2^{\prime} \ell \ell ^{\prime}$ ($Z'$ contribution)}\\
\hline
\multirow{2}{*}{$\mathcal{B}_ {\mu e}\times 10^{-12}$} 
& $\mathcal{S} - \rm I$ & $0.010$ & $0.034$ & $0.063$ & 0.009 & 0.029 & 0.054 \\
\cline{2-8}
& $\mathcal{S} - \rm II$ & 0.044 & 0.141 & 0.262 & 0.037 & 0.117 & 0.218 \\
\hline
\multirow{2}{*}{$\mathcal{B}_ {\tau \mu}\times 10^{-9}$} 
& $\mathcal{S} - \rm I$ & 0.420 & 1.316 & 2.455 & 0.365 & 1.144 & 2.133\\
\cline{2-8}
& $\mathcal{S} - \rm II$ & 0.630 & 2.003 & 3.727  & 0.549 & 1.743 & 3.243\\
\hline
\multirow{2}{*}{$\mathcal{B}_ {\tau e}\times 10^{-9}$} 
& $\mathcal{S} - \rm I$ & $0.525$ & $1.655$ & $3.063$ & 0.457 & 1.440 &  2.666\\
\cline{2-8}
& $\mathcal{S} - \rm II$ & $0.823$ & $2.609$ & $4.851$ & 0.716 & 2.271 & 4.222 \\
\hline
\multirow{2}{*}{$\mathcal{A}_{FB}^{\mu e}$} 
& $\mathcal{S} - \rm I$ & $-0.009$ & $-0.009$ & $-0.009$ & -0.010 & -0.010 & -0.010 \\
\cline{2-8}
& $\mathcal{S} - \rm II$ & -0.120 & -0.120 & -0.120 & -0.123 & -0.123 & -0.123  \\
\hline
\multirow{2}{*}{$\mathcal{A}_{FB}^{\tau \mu}$} 
& $\mathcal{S} - \rm I$ & 0.231 & 0.231 & 0.231 & 0.233 & 0.233 & 0.233 \\
\cline{2-8}
& $\mathcal{S} - \rm II$ & 0.122 & 0.122 & 0.122 & 0.122 & 0.122 & 0.122\\
\hline
\multirow{2}{*}{$\mathcal{A}_{FB}^{\tau e}$} 
& $\mathcal{S} - \rm I$ & 0.247 & 0.247 & 0.247 & 0.248 & 0.248 & 0.248 \\
\cline{2-8}
& $\mathcal{S} - \rm II$ & 0.128 & 0.128 & 0.128  & 0.129 & 0.129 & 0.129 \\
\hline
\hline
\end{tabular}
}
\caption{Upper limit values of $ B \to (K_2^*, f_2^{\prime}) \ell \ell ^{\prime}$ processes in $Z^{\prime}$ model}
\label{tab_BtoK2sf2p}
\end{table}

\begin{figure}[htb]
\centering
\includegraphics[scale=0.57]{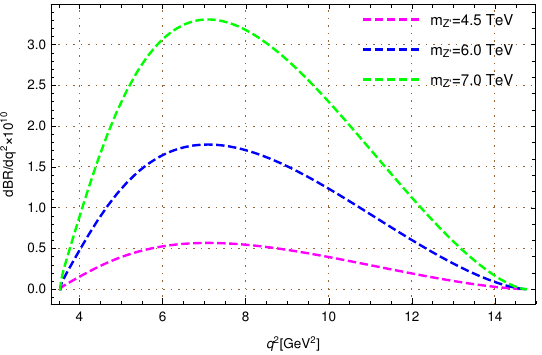}
\quad
\includegraphics[scale=0.57]{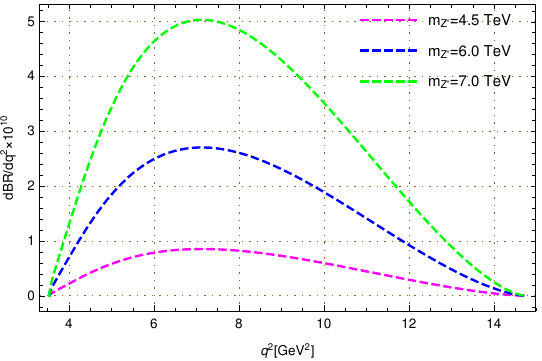}
\quad
\includegraphics[scale=0.57]{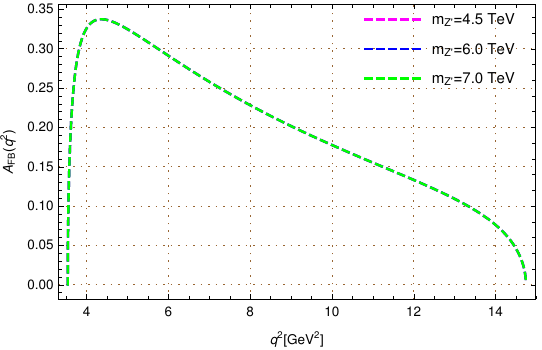} 
\quad
\includegraphics[scale=0.57]{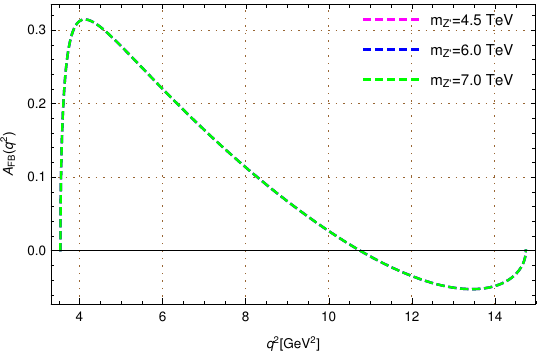}
\caption{The $q^2$ variation of $\mathcal{B}$ and $ A_{FB}$ of $B\to f_2^{\prime} \tau \mu$ channel in the scenario - $\rm I$ (left panel) and scenario - $\rm II$ (right panel).}{\label{Btof2ptaumu}}
\end{figure}

\begin{figure}[htb]
\centering
\includegraphics[scale=0.57]{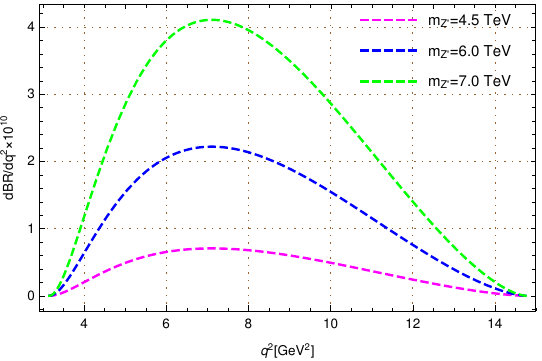}
\quad
\includegraphics[scale=0.57]{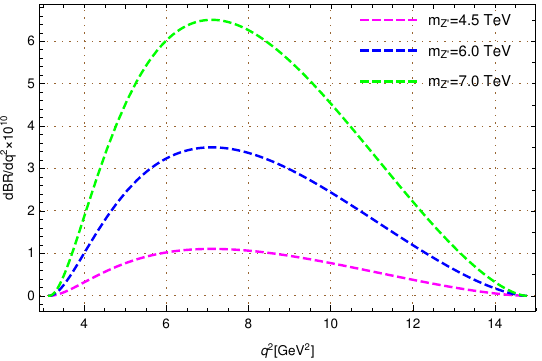}
\quad
\includegraphics[scale=0.57]{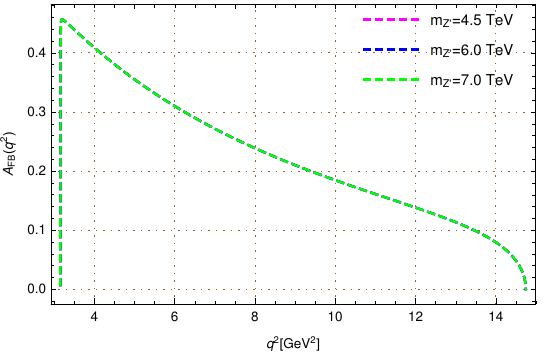} 
\quad
\includegraphics[scale=0.57]{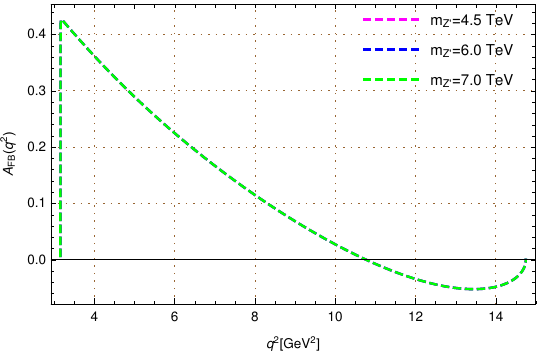}
\caption{ The $q^2$ dependency of the branching ratio and forward-backward asymmetry of $B\to f_2^{\prime} \tau e$ process. $\mathcal{S}- {\rm I}$  and $\mathcal{S}- {\rm II}$, respectively shown in left and right panel.}{\label{Btof2ptaue}}
\end{figure}

\begin{figure}[htb]
\centering
\includegraphics[scale=0.57]{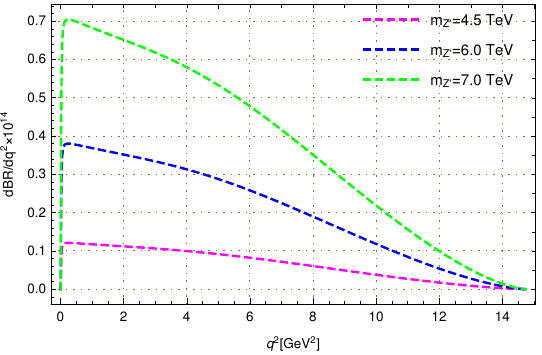}
\quad
\includegraphics[scale=0.57]{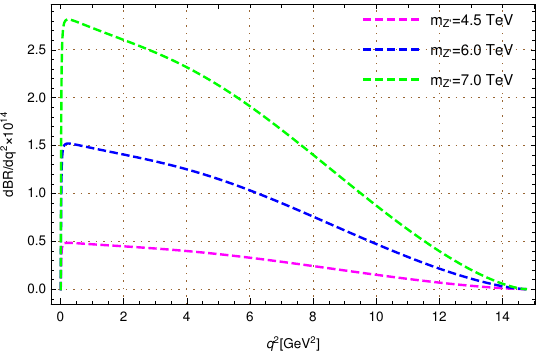}
\quad
\includegraphics[scale=0.57]{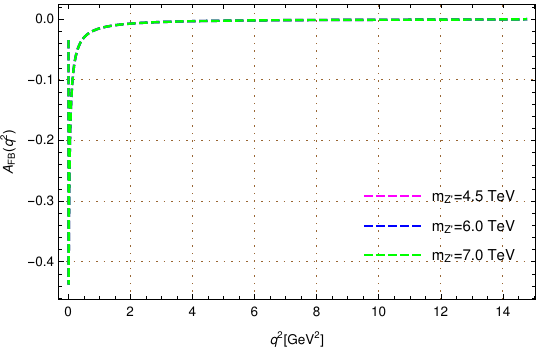}
\quad
\includegraphics[scale=0.57]{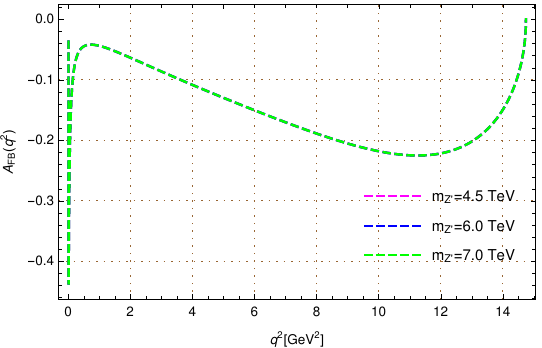}
\caption{Variation of the branching ratio ($\mathcal{B}$) and forward-backward asymmetry ($ A_{FB}$) of $B\to f_2^{\prime} \mu e$ process. Left panel (right panel) indicates $\mathcal{S}- {\rm I}$ ($\mathcal{S}- {\rm II}$).}{\label{Btof2pmue}}
\end{figure}

In Fig.~\ref{BtoK2staumu} and \ref{BtoK2staue}, we analyse the variation of the branching ratio and forward-backward asymmetry of $B\to K_2^* \tau \mu$ and $B\to K_2^* \tau e$ processes with respect to $q^2$, respectively. In both of the figures, the left panel corresponds to the scenario - $\rm I$ whereas the right one indicates to scenario - $\rm II$. The former observable $d\mathcal{B}/dq^2$ contributes distinguishable contributions with higher values in the mid $q^2$ regime with the $m_{Z^{\prime}}$ values. In scenario - $\rm I$ presented in the left panel, the later one has an indistinguishable significant contribution with no zero- crossing point while it allows the same at $q^2 \simeq 10.7$ $\rm GeV^2$ in scenario - $\rm II$  in the presence of $Z^{\prime}$ model. However, there is no change in the contribution for all $m_{Z^{\prime}}$ entries. These are shown in the bottom - left panel of Fig.~\ref{BtoK2staumu} and \ref{BtoK2staue}, respectively. Similarly, the analysis of $B\to K_2^* \mu e$ process is shown in Fig.~\ref{BtoK2smue}. The branching fraction starts from higher values at $q^2=0$ and then reduces to zero in the $ \mu e$ final state. However in the forward backward asymmetry observable, the presence of new physics remains constant at $q^2 \simeq 0$ in scenario - $\rm I$ whereas the the scenario - $\rm II$ indicates significant variations with respect to $q^2$. In both the scenarios, no zero-crossing point has been obsereved. 
Similar to the $B\to K_2^* \ell \ell^{\prime}$ process, we also probe another $B\to T \ell \ell ^{\prime}$ process where $T= f_2^{\prime}$ and $\ell , \ell ^{\prime} = e, \mu, \tau$. With the non-universal $Z^{\prime}$ NP coupling, one can obtain the significant contributions of the branching ratio which are higher than the $B \to K_2^*$ channel. We also investigate the $A_{FB}(q^2)$ observable and obtain quite similar results as compared to previous channel $B\to K_2^* \tau \mu$ and $B\to K_2^* \tau e$.
One can view the plots that have been shown in Fig.~\ref{Btof2ptaumu} and Fig.~\ref{Btof2ptaue} for $B\to f_2 ^ {\prime} \tau \mu$ and $B\to f_2 ^ {\prime} \tau e$ processes, respectively. We also obtain the similar results but with different contributions as the masses and form factors change in $B \to f_2 ^{\prime} \ell \ell ^{\prime}$ process accordingly. Here also we investigate for three $m_{Z^{\prime}} = 4.5, 6.0$ and $7.0$ (in the units of $\rm GeV^2$) values. Similarly in Fig.~\ref{Btof2pmue}, we study the branching ratio and the forward-backward asymmetry of $B \to f_2 ^{\prime} \mu e$ channel and obtain similar results. The top-left  and top-right panel shows the branching ratio whereas the bottom-left and bottom-right panels depict the $A_{FB}(q^2)$ observable in the scenario - $\rm I$ and scenario - $\rm II$, respectively.
We also report the theoretical estimations of the given observables of both $B \to K_2 ^* \ell \ell ^{\prime}$ and $B \to f_2 ^{\prime} \ell \ell ^{\prime}$ processes in Table.~\ref{tab_BtoK2sf2p}. In respect to the scenario - $\rm I$ and $\rm II$, the numerical values of the observables of these LFV decays, presented in the allowed $q^2$ region, differ in the presence of non-universal $Z^{\prime}$ model .

\newpage
\subsection{Lepton non-universality observables}
\begin{figure}[htb]
\centering
\includegraphics[scale=0.57]{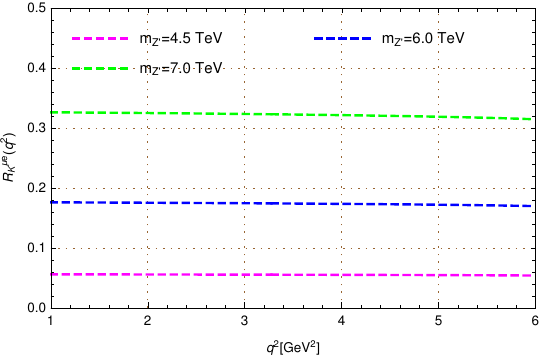}
\quad
\includegraphics[scale=0.57]{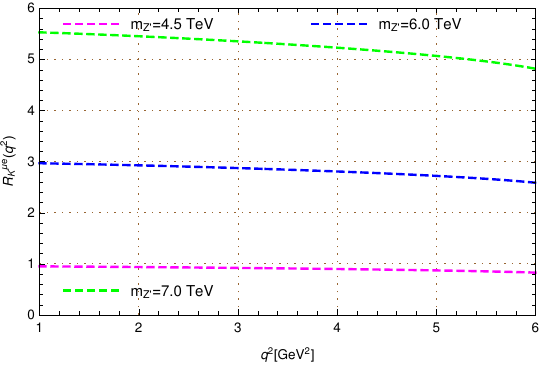}
\quad
\includegraphics[scale=0.57]{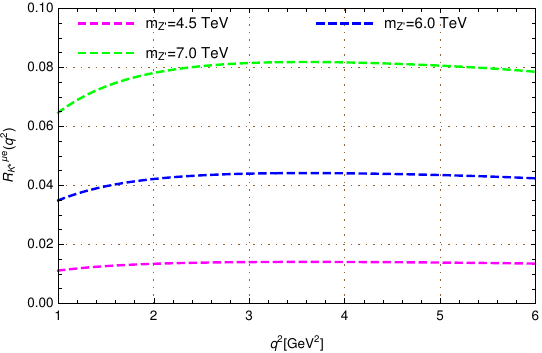}
\quad
\includegraphics[scale=0.57]{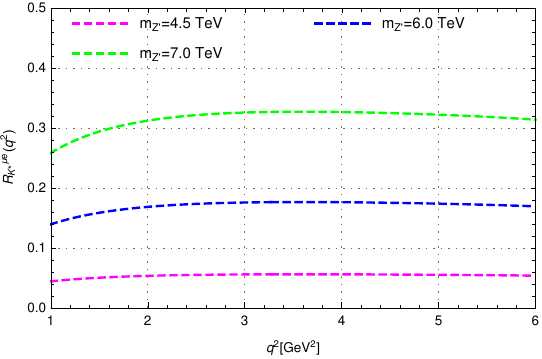}
\quad
\includegraphics[scale=0.57]{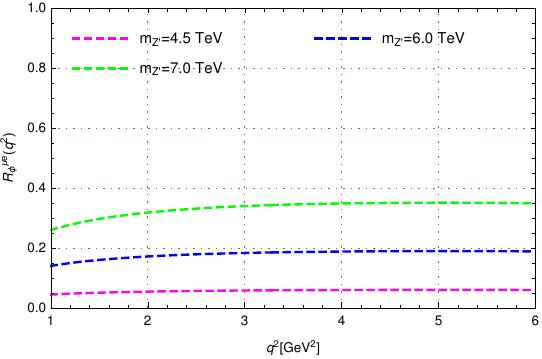}
\quad
\includegraphics[scale=0.57]{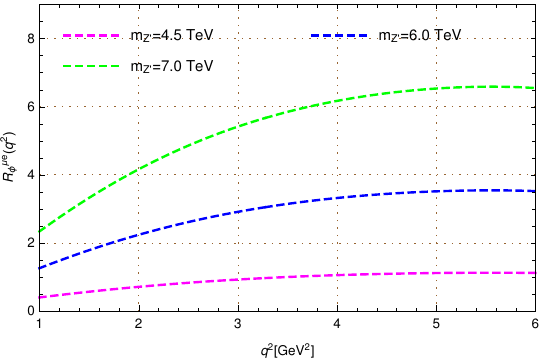}
\quad
\includegraphics[scale=0.57]{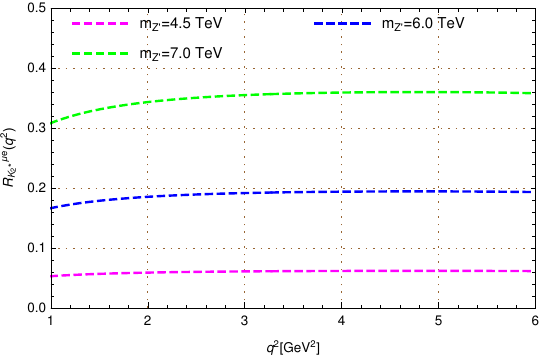}
\quad
\includegraphics[scale=0.57]{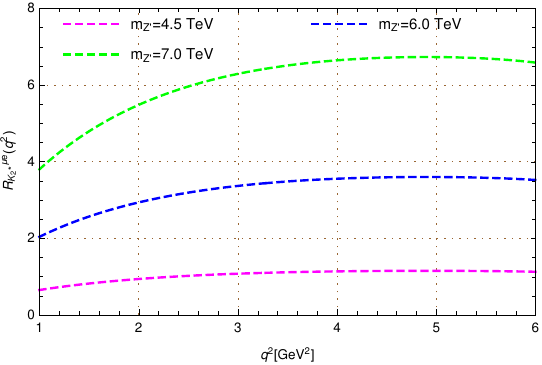}
\quad
\includegraphics[scale=0.57]{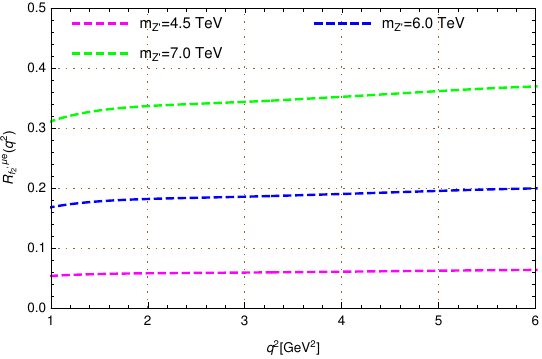}
\quad
\includegraphics[scale=0.57]{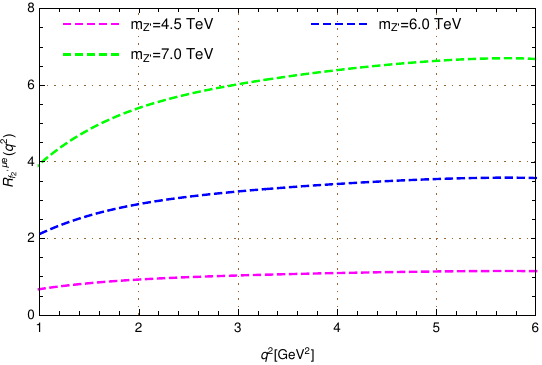}
\caption{The variation of the LNU observables $\mathcal{R}^{\mu e}_{K, K^*, \phi, K_2^*, f_2^{\prime}}$ in the two scenarios: $\mathcal{S}$ - $\rm I$ (left panel) and $\mathcal{S}$ - $\rm II$ (right panel). }{\label{LNUO}}
\end{figure}

Analogous to the clean observable $R_K$ and $R_{K^*}$, we present the behavior of the LNU observable for the exclusive LFV decays given in Eq.~(\ref{LNU}). In the left-panel of  Fig.~\ref{LNUO}, we depict the $q^2$ variations of the LNU observables $\mathcal{R}^{\mu e}_K$, $\mathcal{R}^{\mu e}_{ K^*}$, $\mathcal{R}^{\mu e}_ \phi$, $\mathcal{R}^{\mu e}_{K_2^*}$ and $\mathcal{R}^{\mu e}_{ f_2^{\prime}}$ in scenario - $\rm I$ whereas the right-panel displays the scenario - $\rm II$ in the $q^2 \in [1.0,6.0]$ $\rm GeV^2$ compatible with LHCb measurements. One can visualise that the LNU observable remains constant for different $m_{Z^{\prime}}$ values. All the LNU observables $\mathcal{R}^{\mu e}_{(K, K^*, \phi, K_2^*, f_2^{\prime})}$ are shown with $\mathcal{O} (10^{-6})$ in the given figure. Here the magenta, blue and green line contributions are involved with $m_{Z^{\prime}} = 4.5, 6.0$ and 7.0 $\rm TeV$, respectively. The region $1.0 \leq q^2 \leq  6.0$ behavior says, all the discussed observables show significant contributions with almost a constant value of less than 1. However, the other LNU observables corresponding to $\tau (e, \mu)$ channels couldn't  display the constant values in the given regime. Therefore we haven't considered them in our analysis.  The numerical estimations of all the $\mathcal{R}$ observables are shown in Table.~\ref{tab_R}.

\begin{table}[htp]
\centering
\scalebox{0.9}{
\begin{tabular}{|c|c|c|c|c|c|c|}
\hline
Observable & & $m_{Z^{\prime}}$=4.5 TeV & $m_{Z^{\prime}}$=6.0 TeV & $m_{Z^{\prime}}$=7.0 TeV \\
\hline
\hline
\multicolumn{5}{|c|}{$\mathcal{R}^{\mu e}_{K, K^*, \phi, K_2^*, f_2^{\prime}} \times 10^{-6}$ ($Z'$ contribution)}\\
\hline
\hline
\multirow{2}{*}{$\mathcal{R}^{\mu e}_{K} \times 10^{6}$} 
& $\mathcal{S} - \rm I$ & $0.276$ & $0.869$ & $1.611$  \\
\cline{2-5}
& $\mathcal{S} - \rm II$ & 4.520 & 14.121 & 26.305 \\

\hline $\mathcal{R}^{\mu e}_{K^*}\times 10^{6}$ 
& $\mathcal{S} - \rm I$ & $0.067$ & $0.213$ & $0.395$   \\
\cline{2-5}
& $\mathcal{S} - \rm II$ & $0.271$ & $0.854$ & $1.583$   \\
\hline
\multirow{2}{*}{$\mathcal{R}^{\mu e}_{\phi^*}\times 10^{6}$ } 
& $\mathcal{S} - \rm I$ & $0.287$ & $0.902$ & $1.671$  \\
\cline{2-5}
& $\mathcal{S} - \rm II$ & 4.601 & 14.409 & 26.905 \\
\hline
\multirow{2}{*}{$\mathcal{R}^{\mu e}_{K_2^*}\times 10^{6}$ } 
& $\mathcal{S} - \rm I$ & $0.301$ & $0.948$ & $1.757$  \\
\cline{2-5}
& $\mathcal{S} - \rm II$ & 5.226 & 16.353 & 30.557 \\
\hline
\multirow{2}{*}{$\mathcal{R}^{\mu e}_{f_2^{\prime}}\times 10^{6}$ } 
& $\mathcal{S} - \rm I$ & $0.298$ & $0.939$ & $1.740$  \\
\cline{2-5}
& $\mathcal{S} - \rm II$ & 5.120 & 16.022 & 29.940 \\
\hline
\hline
\end{tabular}
}
\caption{Estimated upper limit values of $\mathcal{R}^{\mu e}_{K, K^*, \phi, K_2^*, f_2^{\prime}}$  LNU observables}
\label{tab_R}
\end{table}
\section{Conclusion}\label{CON}
In this work, we have investigated the flavor violating (semi)leptonic $ B_s \to \ell \ell^{\prime}$, $B_{(s)} \to (K^{(*)}, \phi, f_2^{\prime}$, $K_2^*) \ell \ell ^{\prime}$ channels induced by $b\to s \ell \ell ^{\prime}$ neutral current transition in the presence of non-universal $Z^{\prime}$ model. These decays are extremely rare in the SM because a tiny neutrino mass occur at the loop level. However, the extension of Abelian gauge group $U(1)^{\prime}$ to the SM induces tree level contribution in presence of non-universal $Z^{\prime}$ vector boson. We consider the NP couplings from the branching fractions of  $B_s \to \ell \ell$, $B_s \to \phi \ell \ell$ and the angular observable $P_5^{\prime}$ in the $B \to K^* \ell \ell$ processes with the naive $\chi ^2$ analysis. Using such couplings, we mainly analyse the variation of the branching fractions, forward-backward asymmetries, polarisation asymmetries of all the associated semi(leptonic) $ B_s \to \ell \ell^{\prime}$, $B_{(s)} \to (K^{(*)}, \phi, f_2^{\prime}$, $K_2^*) \ell \ell ^{\prime}$ decay channels in the presence of non-universal $Z^{\prime}$ boson. We also compute the theoretical values of all the observables.  To inspect the presence of lepton non-universality, we construct and analyse the observables $\mathcal{R}^{\mu e}_{K, K^*, \phi, K_2^*, f_2^{\prime}}$ in the $q^2 \in [1.0,6.0]$ regime which are compatible with the LHCb measurement. We obtain that the $q^2$ variations of the observables have distinguished contributions in the presence of NP couplings and three different $m_{Z^{\prime}}$ values. Additionally, the theoretically estimated values are sizeable and have definite contributions. However, these decay channels could be further analysed in upcoming LHCb and B-factories with large number of events which could lead to the origin of univocal signal of new physics.
\acknowledgments 
LN and RD would like to acknowledge DST INSPIRE fellowship programme for financial support.

\appendix
\section{The $\mathcal{\phi}(q^2)$ parameters in $B \to K \ell \ell ^{\prime}$ process} \label{BtoKparameters}
The $\mathcal{\phi} (q^2)$ parameters used in the $B \to K \ell \ell ^{\prime}$ are gievn as below
\begin{align}
\label{eq:C910}
\varphi_{7}(q^2) &=  \frac{2 m_b^2|f_T(q^2)|^2}{(m_B+m_K)^2} \lambda(m_B,m_K,\sqrt{q^2})\left[1-\frac{(m_1-m_2)^2}{q^2}-\frac{\lambda(\sqrt{q^2},m_1,m_2)}{3 q^4}\right], \nn \\
\varphi_{9(10)}(q^2)&=\frac{1}{2}|f_0(q^2)|^2(m_1\mp m_2)^2 \frac{(m_B^2-m_K^2)^2}{q^2} \left[1-\frac{(m_1\pm m_2)^2}{q^2}\right] \nn \\
&+\frac{1}{2}|f_+(q^2)|^2 \lambda(m_B,m_K,\sqrt{q^2})\left[1-\frac{(m_1\mp m_2)^2}{q^2}-\frac{\lambda(\sqrt{q^2},m_1,m_2)}{3 q^4}\right] \nn ,\\
\varphi_{79}(q^2)&=\frac{2 m_b f_+(q^2)f_T(q^2)}{m_B+m_K} \lambda(m_B,m_K,\sqrt{q^2})\left[ 1-\frac{(m_1-m_2)^2}{q^2}-\frac{\lambda(\sqrt{q^2},m_1,m_2)}{3 q^4}\right],\nn \\
\varphi_{S (P)}(q^2)&=\frac{q^2 |f_0(q^2)|^2}{2(m_b-m_s)^2}(m_B^2-m_K^2)^2 \left[1-\frac{(m_1\pm m_2)^2}{q^2}\right], \nn \\
\varphi_{10P (9S)}(q^2)&=\frac{|f_0(q^2)|^2}{m_b-m_s}(m_1\pm m_2)(m_B^2-m_K^2)^2\left[1-\frac{(m_1 \mp m_2)^2}{q^2}\right].
\end{align}
In  the function $\varphi_{a(b)}(q^2)$ the upper sign represents $\varphi_a(q^2)$ whereas the lower one to $\varphi_b(q^2)$.

\section{The angular coefficient parameters in $B \to (K^*, \phi) \ell \ell ^{\prime}$ processes} \label{BtoKstarparameters}
The parameters $I_i^j(q^2)~(i=1,2; j= c,s)$ are the $q^2$- dependent angular coefficients. These include the transversity amplitudes $ A_{\perp,\parallel,0,t}^{L(R)}(q^2)$ and are given as follows: 
\begin{align}
A_{\perp}^{L(R)} &= {\cal N}_{K^\ast} \sqrt{2} \lambda_B^{1/2}\left[[(C_9+C_9')\mp(C_{10}+C_{10}')]\frac{V(q^2)}{m_B+m_{K^\ast}}+\frac{2 m_b}{q^2}(C_7+C_7') T_1(q^2)\right],\nn \\
A_{\parallel}^{L(R)} &=  -{\cal N}_{K^\ast} \sqrt{2}(m_B^2-m_{K^\ast}^2)\left[[(C_9-C_9')\mp (C_{10}-C_{10}')]\frac{A_1(q^2)}{m_B-m_{K^\ast}}+\frac{2 m_b}{q^2}(C_7-C_7')T_2(q^2)\right],\nn  
\end{align}
\begin{align}
A_0^{L(R)}&=-\frac{{\cal N}_{K^\ast}}{2 m_{K^\ast} \sqrt{q^2}}\Big{\lbrace} 2 m_b (C_7-C_7')\left[(m_B^2+3m_{K^\ast}^2-q^2)T_2(q^2)-\frac{ \lambda_B T_3(q^2)}{m_B^2-m_{K^\ast}^2} \right]\nn \\
&+[(C_9-C_9')\mp(C_{10}-C_{10}')]\cdot\left[ (m_B^2-m_{K^\ast}^2-q^2)(m_B+m_{K^\ast})A_1(q^2)-\frac{ \lambda_B A_2(q^2)}{m_B+m_{K^\ast}}\right] \Big{\rbrace} \nonumber\\[0.7 em]
\label{eq:helicityamplitudest}
A_{t}^{L(R)} &=  -{\cal N}_{K^\ast} \frac{\lambda_B^{1/2}}{\sqrt{q^2}}\left[(C_9-C_{9}') \mp (C_{10}-C_{10}') +\frac{q^2}{m_b+m_s}\left(\frac{C_S-C_S'}{m_1-m_2}\mp \frac{C_P-C_P'}{m_1+m_2}\right)\right] A_0(q^2)
\end{align}
with 
\begin{equation}
{\cal N}_{K^\ast}=V_{tb}V_{ts}^\ast\left[ \frac{\tau_{B_d} G_F^2 \alpha^2}{3 \times 2^{10} \pi^5 m_B^3} \lambda_B^{1/2} \lambda_q^{1/2} \right]^{1/2}.
\end{equation}

The kinematic factors are
$\lambda_B=\lambda(m_B,m_{K^\ast},\sqrt{q^2})$ and $\lambda_q=\lambda(m_1,m_2,\sqrt{q^2})$ where the corresponding formula are given in Eq.~(\ref{knfactor}). 

The angular coefficients $I_{1-9}(q^2)$ in terms of the transversity amplitudes~(\ref{eq:helicityamplitudest}) are given as
\begin{align}
\label{eq:angular}
I_1^s(q^2) &=\biggl[|A_{\perp}^L|^2+|A_{\parallel}^L|^2+ (L\to R) \biggr]\frac{\lambda_q +2 [q^4-(m_1^2-m_2^2)^2]}{4 q^4}+\frac{4 m_1 m_2}{q^2}\mathrm{Re}\left(A_{\parallel}^L A_{\parallel}^{R\ast}+A_{\perp}^L A_{\perp}^{R\ast}\right), \nn \\
I_1^c(q^2) &= \bigl[|A_0^L|^2+|A_0^R|^2 \bigr]\frac{q^4-(m_1^2-m_2^2)^2}{q^4}+\frac{8 m_1 m_2}{q^2} \mathrm{Re}(A_0^L A_0^{R\ast}-A_t^L A_t^{R\ast}) \nn\\
&\hspace{3.5cm}-2\frac{(m_1^2-m_2^2)^2-q^2 (m_1^2+m_2^2)}{q^4}\bigl(|A_t^L|^2+|A_t^R|^2\bigr),\nonumber\\
I_2^s(q^2) &= \frac{\lambda_q}{4 q^4}[|A_\perp^L|^2+|A_\parallel^L|^2+(L\to R)], \nn  \\
I_2^c(q^2) &= - \frac{\lambda_q}{q^4}(|A_0^L|^2+|A_0^R|^2), \nn \\
I_3(q^2) &= \frac{\lambda_q}{2 q^4} [|A_\perp^L|^2-|A_\parallel^L|^2+(L\to R)],\nn \\
I_4(q^2) &= - \frac{\lambda_q}{\sqrt{2} q^4} \mathrm{Re}(A_\parallel^L A_0^{L\ast}+(L\to R)],\nn \\
I_5(q^2) &= \frac{\sqrt{2}\lambda_q^{1/2}}{q^2} \left[ \mathrm{Re}(A_0^L A_\perp^{L\ast}-(L\to R)) -\frac{m_1^2-m_2^2}{q^2} \mathrm{Re}(A_t^L A_\parallel^{L\ast}+(L\to R))\right], \nn \\
I_6^s(q^2) &=- \frac{2 \lambda_q^{1/2}}{q^2}[\mathrm{Re}(A_\parallel^L A_\perp^{L\ast}-(L\to R))],\nn  \\
I_6^c(q^2) &= - \frac{4\lambda_q^{1/2}}{q^2}\frac{m_1^2-m_2^2}{q^2} \mathrm{Re}(A_0^L A_t^{L\ast}+(L\to R)),\nn \\
I_7(q^2) &= - \frac{\sqrt{2}\lambda_q^{1/2}}{q^2} \left[ \mathrm{Im}(A_0^L A_\parallel^{L\ast}-(L\to R))+ \frac{m_1^2-m_2^2}{q^2} \mathrm{Im}(A_\perp^{L}A_t^{L\ast} +(L\to R))\right], \nn \\
I_8(q^2) &= \frac{\lambda_q}{\sqrt{2}q^4}\mathrm{Im}(A_0^{L}A_\perp^{L\ast} +(L\to R)), \nn \\
I_9(q^2) &=- \frac{\lambda_q}{q^4}\mathrm{Im}(A_\perp^L A_\parallel^{L\ast} +A_\perp^R A_\parallel^{R\ast}  ),
\end{align}

\section{Required parameters of $ B \to T(K^*_2, f_2 ^{\prime}) \ell \ell ^{\prime}$}
\label{appenb}
The $q^2$ parameters of $B \to T{\ell \ell ^{\prime}}$ are given as follows
\begin{eqnarray}
A(q^2) &=& \frac{3}{4}\left\lbrace\frac{1}{4}\left[\left(1+\frac{\mpl^2}{q^2}\right)\bmi^2 +\left(1+\frac{\mm^2}{q^2}\right)\bpl^2\right]\left(|A_L^{\parallel}|^2+|A_L^{\perp}|^2+(L\to R)\right)\right.\nn\\
&&+\frac{1}{2}\left(\bpl^2 +\bmi^2\right)\left(|A_L^{0}|^2+|A_R^{0}|^2\right)\nn\\
&&+\frac{4m_1 m_2}{q^2} \re\left[ A_R^0 A_L^{0*} +A_R^{\parallel} A_L^{\parallel *}+A_R^{\perp} A_L^{\perp*}-A_L^t A_R^{t*}\right]\nn\\
&&+\frac{1}{2}\left(\bmi^2+\bpl^2-2\bmi^2\bpl^2\right)\left(|A_L^t|^2 + |A_R^t|^2\right) +\frac{1}{2} \left(|A_{SP}|^2 \bmi^2 +|A_{S}|^2 \bpl^2\right)\nn\\
&&\left.+\frac{2\mm}{ \sqrt{q^2}}\bpl^2 \re\left[A_S(A_L^t+A_R^t)^*\right]- \frac{2\mpl}{\sqrt{q^2}} \bmi^2 \re\left[A_{SP}(A_L^t-A_R^t)^*\right]\right\rbrace,\\ \label{dist:A}
B(q^2)&=&\frac{3}{2} \bmi \bpl\left\lbrace \re\left[A_{L}^{\perp *} A_L^{\parallel} -(L \to R)\right]+ \frac{m_+m_-}{q^2} \re\left[A_{L}^{0*} A_L^{t} +(L \to R)\right]\right.\nn\\
&& \left. + \frac{m_{+}}{\sqrt{\qsq}} \re\left[A_S^* (A_L^0 +A_R^0) \right]-\frac{\mm} {\sqrt{\qsq}} \re\left[A_{SP}^* (A_L^0 -A_R^0) \right]\right\rbrace,\\
C(q^2)&=&\frac{3}{8}\bpl^2 \bmi^2  \left\lbrace\left(|A_L^{\parallel}|^2+|A_L^{\perp}|^2-2|A_L^{0}|^2\right)+\left(L\to R\right)\right\rbrace \label{dist:C}
 \end{eqnarray}
 Here $m_{\pm}=(m_1 \pm m_2)$, $\beta_{\pm} = \sqrt{1-\frac{(m_{\ell} \pm m_{\ell^{\prime}})^2}{q^2}} $ and the expressions of transversity amplitudes $A$'s are given in Appendix~\ref{Tamp}.

The polarization $\epsilon^{\mu\nu}(n)$ of tensor meson $K_2^\ast$, which has four momentum $(k_0, 0, 0, \vec{k})$, can be written in terms of the spin-1 polarization vectors~\cite{Berger:2000wt}
\begin{eqnarray}
 \epsilon_{\mu\nu}(\pm 2) &=& \epsilon_\mu(\pm 1)\epsilon_\nu(\pm 1),\nn\\
 \epsilon_{\mu\nu}(\pm 1) &=& \frac{1}{\sqrt{2}}\left[\epsilon_\nu(\pm)\epsilon_\nu(0) + \epsilon_\nu(\pm)\epsilon_\mu(0) \right],\nn\\
 \epsilon_{\mu\nu}(0) &=& \frac{1}{\sqrt{6}}\left[\epsilon_\mu(+) \epsilon_\nu(-) + \epsilon_\nu(+) \epsilon_\mu(-) \right]+ \sqrt{\frac{2}{3}}\epsilon_\mu(0)\epsilon_\nu(0) ,
 \end{eqnarray} 
where the spin-1 polarization vectors are defined as
\begin{equation}
\epsilon_\mu(0) = \frac{1}{m_{K_2^\ast}}\left(\vec{k}_z,0,0,k_0\right)\, ,\quad \epsilon_\mu(\pm) = \frac{1}{\sqrt{2}}\left(0,1,\pm i, 0\right)\
\end{equation}

In the  study of $B\to T (K_2^*, f_2 ^{\prime})\ell_1\ell_2$ decay channel, it has two leptons in the final state so the $n=\pm 2$ helicity states of the $K_2^\ast$
is not realized. So a new polarization vector is introduced~\cite{Wang:2010tz}
\begin{equation}
\epsilon_{T\mu}(h) = \frac{\epsilon_{\mu\nu}p^\nu}{m_B}\, 
\end{equation}
 The explicit expressions of polarization vectors are
\begin{eqnarray}
\epsilon_{T\mu}(\pm 1) &=& \frac{1}{m_B}\frac{1}{\sqrt{2}}\epsilon(0).p  \epsilon_\mu(\pm) = \frac{\sqrt{\lambda}}{\sqrt{8}m_B m_{K^*_2}} \epsilon_\mu(\pm), \\
\epsilon_{T\mu}(0) &=& \frac{1}{m_B}\sqrt{\frac{2}{3}}\epsilon(0).p \epsilon_\mu(0) = \frac{\sqrt{\lambda}}{\sqrt{6}m_B m_{K^*_2}} \epsilon_\mu(0),
\end{eqnarray}
where $\lambda(m^2_B,m^2_{K^*_2},q^2) = m^4_B + m^4_{K^*_2} + q^4 -2(m^2_B m^2_{K^*_2}+m^2_Bq^2+m^2_{K^*_2}q^2)$ is the usual Kallen function.
On the other hand, the virtual gauge boson can have three types of polarization states, longitudinal, transverse and time-like, which have following components
\begin{equation}
\epsilon^\mu_V(0) = \frac{1}{\sqrt{q^2}}(-|\vec{q_z}|,0,0,-q_0)\, ,\quad \epsilon^\mu_V(\pm) = \frac{1}{\sqrt{2}}(0,1,\pm i, 0)\ ,\quad
\epsilon^\mu_V(t) = \frac{1}{\sqrt{q^2}}(q_0,0,0,q_z)\ 
\end{equation}
where $q^\mu=(q_0,0,0,q_z)$ is four momentum of gauge boson.

\subsection{Transversity Amplitudes}
\label{Tamp}
The vector and axial-vector transversity amplitudes can be expressed as
 \begin{eqnarray}
A_{0L,R} &=& N  \frac{\sqrt{\lambda}}{\sqrt6 m_Bm_{K_2^*}}\frac{1}{2m_{K^*_2}\sqrt {q^2}}\left[ (C_{V-}\mp C_{A-})
\left[(m_B^2-m_{K^*_2}^2-q^2)(m_B+m_{K^*_2})A_1 -\frac{\lambda}{m_B+m_{K^*_2}}A_2 \right] \right], \nonumber\\
A_{\perp L,R} &=& -\sqrt{2} N \frac{\sqrt{\lambda}}{\sqrt8m_Bm_{K_2^*}}\left[(C_{V+}\mp C_{A+})
 \frac{\sqrt \lambda V}{m_B+m_{K^*_2}}\right],   \nonumber \\
A_{\parallel L,R} &=& \sqrt{2} N\frac{\sqrt{\lambda}}{\sqrt{8} m_B m_{K_2^*}} \left[(C_{V-}\mp C_{A-}) (m_B+m_{K^*_2}) A_1\right], \nonumber \\
A_{Lt} &=&N\frac{\sqrt{\lambda}}{\sqrt{q^2}\sqrt{6}m_B m_{K^\ast_2}}\left[ \sqrt{\lambda}(C_{V-}-C_{A-}) A_0\right],\nonumber \\
A_{Rt} &=&N\frac{\sqrt{\lambda}}{\sqrt{q^2}\sqrt{6}m_B m_{K^\ast_2}}\left[ \sqrt{\lambda}(C_{V-}+C_{A-}) A_0\right],
\end{eqnarray}
where $C_{V\pm} = (C_{V}  \pm C_{V}^\prime)$, and $C_{A\pm} = (C_{A} \pm C_{A}^\prime)$.
The transversity amplitudes for scalar, pseudoscalar  interactions can be written as 
\begin{eqnarray}
  A_S  &=& 2 N \frac{\sqrt{\lambda}}{\sqrt{6}m_B m_{K^\ast_2}}\left [\sqrt{\lambda}(C_{S} - C_{S'}) A_0 \right],\nonumber \\
  A_{SP} &=& 2 N\frac{\sqrt{\lambda}}{\sqrt{6}m_B m_{K^\ast_2}}\left[\sqrt{\lambda}(C_P-C_{P'})A_0\right].
\end{eqnarray}
The normalization constant $N$ is given by
\begin{equation}
    N = \left[ \frac{G_F^2\alpha_e^2}{3\cdot 2^{10}\pi^5 m_B^3}|V_{tb}V_{ts}^\ast|^2 q^2\bpl \bmi \lambda(m^2_B,m^2_{K^*_2},q^2)^{1/2} \mathcal{B}(K_2^\ast\to K\pi) \right]^\frac{1}{2}.
\end{equation}
\bibliographystyle{ieeetr}
\bibliography{ALRM}
\end{document}